\documentclass[twocolumn, trackchanges]{aastex63}

\usepackage{array}
\usepackage{booktabs}
\usepackage{multirow}
\usepackage{makecell}
\usepackage{hyperref}
\usepackage[hyphenbreaks]{breakurl}
\usepackage{balance}
\usepackage{amsmath}
\usepackage{enumerate}

\shorttitle{Witnessing the Star Formation quenching in the protocluster core}
\shortauthors{Shi et al.}

\begin{document}



\title{The Emergence of  Brightest Cluster Galaxy in a Protocluster Core at $z=2.24$}

\email{ddshi@pmo.ac.cn, xzzheng@pmo.ac.cn}

\author[0000-0002-4314-5686]{Dong~Dong~Shi}
\affiliation{Purple Mountain Observatory, Chinese Academy of Sciences, 10 Yuan Hua Road, Nanjing, Jiangsu 210023, China}

\author[0000-0002-9373-3865]{Xin~Wang}
\affil{School of Astronomy and Space Sciences, University of Chinese Academy of Sciences (UCAS), Beijing 100049, China}
\affil{National Astronomical Observatories, Chinese Academy of Sciences, Beijing 100101, China}
\affil{Institute for Frontiers in Astronomy and Astrophysics, Beijing Normal University, Beijing 102206, China}

\author[0000-0003-3728-9912]{Xian~Zhong~Zheng}
\affiliation{Purple Mountain Observatory, Chinese Academy of Sciences, 10 Yuan Hua Road, Nanjing, Jiangsu 210023, China}
\affiliation{School of Astronomy and Space Sciences, University of Science and Technology of China, Hefei 230026,China}

\author[0000-0001-8467-6478]{Zheng~Cai}
\affiliation{Department of Astronomy, Tsinghua University, Beijing 100084, China}

\author[0000-0003-3310-0131]{Xiaohui~Fan}
\affiliation{Steward Observatory, University of Arizona, 933 North Cherry Avenue, Tucson, AZ 85721, USA}

\author[0000-0002-1620-0897]{Fuyan~Bian}
\affiliation{European Southern Observatory, Alonso de C{\'o}rdova 3107, Casilla 19001, Vitacura, Santiago 19, Chile}

\author[0000-0002-7064-5424]{Harry~I.~Teplitz}
\affiliation{IPAC, Mail Code 314-6, California Institute of Technology, 1200 E. California Blvd., Pasadena CA, 91125, USA}






\begin{abstract}

We report the detection of a pair of massive quiescent galaxies likely in the process of merging at the center of the spectroscopically confirmed, extremely massive protocluster BOSS1244 at  $z=2.24\pm0.02$.  These galaxies, BOSS1244-QG1  and BOSS1244-QG2, were detected with Hubble Space Telescope (HST) grism slitless spectroscopic observations. These two quiescent galaxies are among the brightest member galaxies with $z=2.223-2.255$ in BOSS1244 and reside at redshifts $z=2.244$ and $z=2.242$, with a half-light radius of $6.76\pm0.50$ and $2.72\pm0.16$\,kpc, respectively. BOSS1244-QG1 and BOSS1244-QG2 are separated by a projected distance of about 70\,physical\,kpc, implying that the two galaxies likely merge to form a massive brightest cluster galaxy (BCG) with size and mass similar to the most massive BCGs in the local Universe. We thus infer that BCG formation through dry major mergers may happen earlier than the full assembly of a cluster core, which broadens our previous understanding of the co-evolution of mature galaxy clusters and BCGs in the nearby Universe. 
Moreover, we find a strong density-star formation relation over a scale of $\sim18$\,co-moving\,Mpc in BOSS1244,  i.e. star formation activity decreases as density increases, implying that the quenching of star formation in BCGs and their progenitors is likely governed by environment-related processes before the virialization of the cluster core.

\end{abstract}



\section{Introduction} \label{sec:intro}

Massive clusters of galaxies are the densest structures in the Universe and contain the most massive galaxies called the Brightest Cluster Galaxies (BCGs). The formation of these BCGs in massive galaxy clusters occurs under extreme conditions, making it an ideal testing ground for theories of galaxy formation in relation to structure formation \citep[e.g.,][]{DeLucia2007}. In recent decades, significant progress has been made in understanding the properties of galaxies in massive galaxy clusters across cosmic time. However, while BCGs in nearby clusters have been extensively studied, the early formation of BCGs in galaxy protoclusters remains poorly understood, despite efforts to combine observations and theory.

In nearby galaxy clusters, the well-established ``morphology-density" relation of member galaxies and the presence a tight red sequence of quiescent galaxies evidence strong environmental effects on galaxy evolution in morphology and star formation \citep{Dressler1980,Bower1992,Goto2003,Postman2005, Blakeslee2006, Patel2009, Mei2009, Muzzin2012, Foltz2015,Foltz2018}.  Compared with their analogues in general fields, massive quiescent galaxies in the clusters were formed earlier  \citep{Thomas2005},  attributing to an acceleration of galaxy evolution and an enhancement of star formation in the early assembling epoch of the clusters \citep{Elbaz2007, Scoville2007, Cooper2010, Patel2011, Quadri2012}. 
At $z>1.5$, it becomes difficult to spectroscopically identify quiescent galaxies with ground-based facilities. Still, pioneering studies reported a number of detections of high-$z$ quiescent galaxies \citep[e.g.,][]{Dunlop1996, Cimatti2004, Daddi2005, Kriek2006,Kriek2009, Gobat2012, Gobat2013, Newman2014, Belli2014, Kriek2015, Belli2017, Belli2019, Strazzullo2015, Glazebrook2017, Forrest2020, Willis2020, DEugenio2021, Kubo2021,McConachie2021}. Only several studies confirmed  quiescent galaxies with Balmer absorption features in the protoclusters at $z=2-3$, and  the presence of a  higher  fraction of quiescent galaxies  in the (near-)virialized clusters \citep{Kubo2021, McConachie2021, Ito2023}.  These suggest that  high-$z$ galaxy (proto)clusters may be in various evolutionary stages and some systems already harbor some galaxies with star formation quenched in the central regions. Whether these quiescent galaxies at the center of galaxy protoclusters are the progenitors of local BCGs remains to be explored.

The properties of nearby BCGs have been widely investigated \citep{Jones1984, Postman1995, Lin2004, VonDerLinden2007, Lauer2014, Zhao2017}.  They are typically quiescent, having a specific star formation rate (sSFR) at the level of $\sim0.001$\,Gyr$^{-1}$ \citep{Fraser-McKelvie2014}.  They are usually located close to the peak of the cluster X-ray emission.  BCGs exhibit a steeper size--luminosity relation (some present extended envelopes with cD morphology) than elliptical galaxies in the field, suggesting that BCGs evolve through dry major mergers of quiescent galaxies \citep{Bernardi2007, Liu2009, Liu2015, vanDokkum2015}. Their unique properties (e.g., large size and large velocity dispersion) and living environments make them differ from the coeval field early-type galaxies.  The active galactic nuclei (AGN) feedback from BCGs are widely proposed to act as an important mechanism to heat the intracluster medium (ICM) and prevent star-formation in BCGs \citep{Fabian2012, Man2018}.

Early theoretical studies proposed that the formation of BCGs is driven by cooling flows in the cores of galaxy clusters and galactic cannibalism through dynamical friction \citep{Ostriker1975, White1976, Richstone1976, Fabian1994}.  However, cosmological simulations suggest  that BCG formation occurs hierarchically through multiple galaxy mergers \citep{DeLucia2007}, while observational studies confirm that some BCGs were already present at $z=1-1.5$ and underwent   rapid early growth rather than hierarchical assembly \citep{Collins2009}.  Additionally,  dissipationless mergers of massive quiescent galaxies at $z\sim 2$ have been identified as a possible pathway for BCG formation \citep{Laporte2013}. Nevertheless, there are limited observational studies contributed to the early formation and star formation quenching of BCGs at high redshifts ($z>2$). Therefore, direct observations of BCGs at high redshifts are crucial for understanding their connection to host clusters, as well as their formation and evolution. The distant protoclusters of galaxies, which are the progenitors of massive galaxy clusters in the present day, are expected to host BCGs in their early formation stages.

Recently, we conducted a study on a protocluster located at $z=2.24$ in the BOSS1244 field. This protocluster was selected through a group of strong Ly$\alpha$ absorbers and QSOs, and further confirmed by the high density of narrowband-selected H$\alpha$ emitters (HAEs) at $z=2.24\pm0.02$ \citep{Cai2016, Zheng2021}. Near-infrared (NIR) spectroscopy of the HAEs revealed two distinct components in the southwest (SW) region of BOSS1244. Notably, the BOSS1244 SW region, which comprises of two substructures, represents the most overdense galaxy protocluster discovered at $z>2$ to date. It is expected to eventually evolve into the most massive Coma-type galaxy cluster at $z\sim0$ \citep{Shi2021}.

Based on the velocity dispersion measured from with the NIR spectroscopy, \cite{Shi2021} estimated the dynamical mass of the BOSS1244 SW region to be $3 \times$10$^{13}\,$M$_{\odot}$ , implying that the core halo of this protocluster is transitioning from the cold mode to the formation stage of shock-heated gas (referred to as the ``hot" mode) \citep{Dekel2006}. This suggests that the protocluster is currently in a maturing stage \citep{Shimakawa2018}. Additionally, the outskirts of the BOSS1244 SW region display an increased density of submillimeter galaxies (SMGs), which were detected by the deep 850\,$\micron$ observations with the James Clerk Maxwell Telescope (JCMT)/Submillimeter Common-User Bolometer Array 2 (SCUBA-2). Interestingly, these SMGs show a notable spatial offset from the density peak of the HAEs \citep{Zhang2022}. This discrepancy indicates that the core of the protocluster in the BOSS1244 SW region may currently be experiencing a lack of intense star formation (starburst) activity. Consequently, this protocluster offers a unique opportunity to observe the transition of starburst galaxies into quiescent galaxies.

\begin{figure*}[!ht]
\centering
\includegraphics[trim=0mm 0mm 0mm 1mm,clip,width=0.9\textwidth]{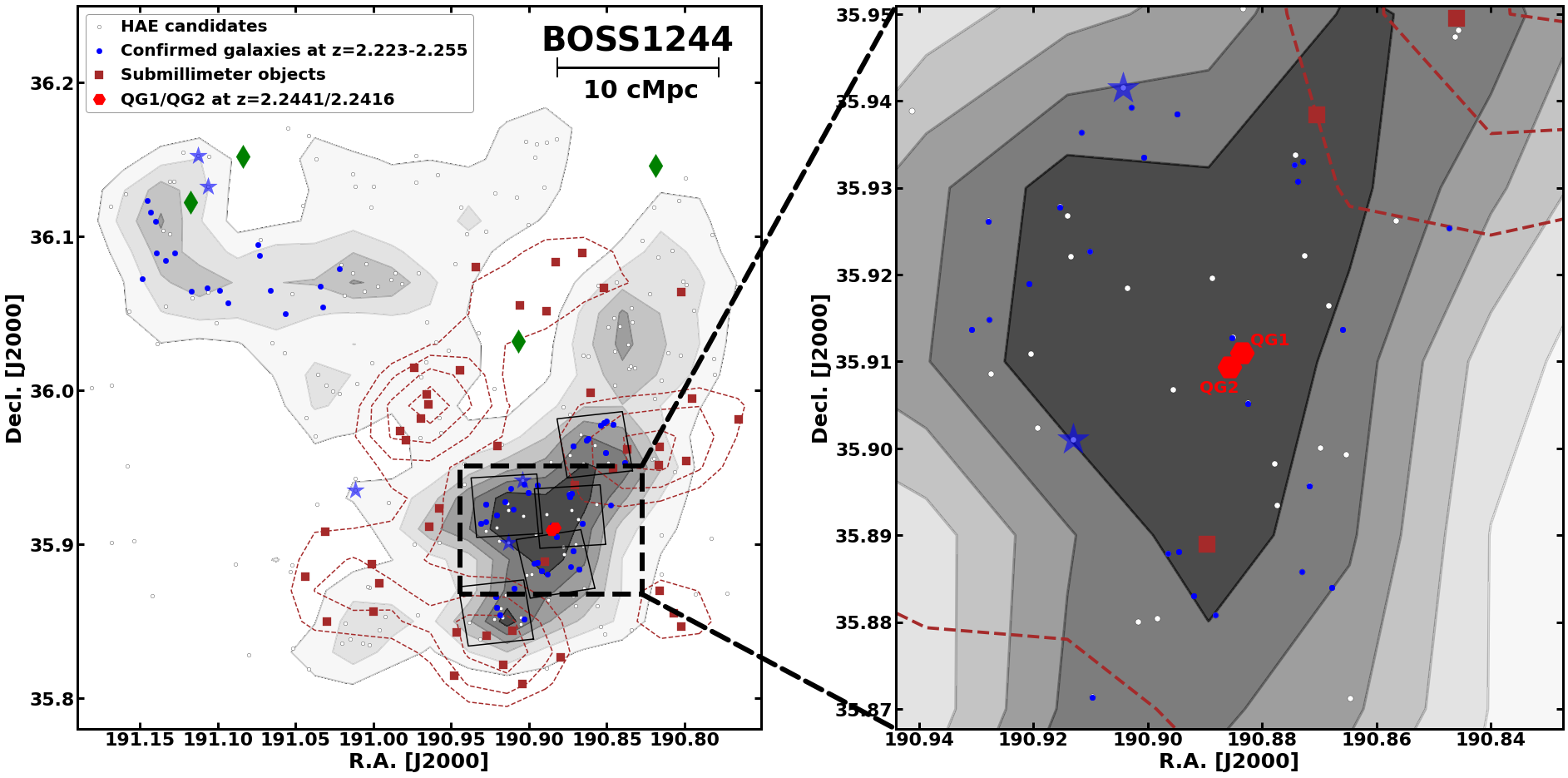}
\caption{Spatial distribution of member galaxies in BOSS1244 at $z=2.24$.  {\bf Left:} Density maps of HAEs (grey solid contour) and SMGs (brown dotted contour) in BOSS1244. The contour levels are adopted from \cite{Zheng2021} and \cite{Zhang2022}. The blue points are the confirmed member star-forming galaxies at $z=2.223-2.255$ from HST grism and ground-based NIR spectroscopy. The white points are the NB-selected HAE candidates and the brown squares are the submillimeter objects from JCMT/SCUBA-2 observations  at 850\,$\micron$.  The blue stars are the SDSS/BOSS quasars at $z = 2.24 \pm 0.02$ and the green diamonds show groups of Ly$\alpha$ absorption systems. The red hexagons are BOSS1244-QG1 at $z=2.2441$ and BOSS1244-QG2 at $z=2.2416$. Five black solid boxes  in the density peak of BOSS1244 show the coverage of  HST observations. The size of black dashed box is $3\farcm5 \times 2\farcm5$. {\bf Right:}  Zoomed-in on the left black dashed box. The positions of two massive quiescent galaxies are marked by  ``QG1'' and ``QG2''.} 
\label{fig:fig1}
\end{figure*}

We carried out the Hubble Space Telescope Wide Field Camera 3 (HST/WFC3) grism slitless spectroscopic survey to obtain followup slitless spectroscopy of the protocluster BOSS1244 SW  region \citep{Wang2021}. Compared with ground-based telescopes, HST/WFC3 in its grism has high throughput and sensitivity, and the observations are free from the atmospheric absorption and high sky background nor biased surveys from photometric pre-selection. Combing the relatively low spectral resolution (e.g., G141, $R=130$) and limited filed-of-view (FOV, $\sim2\farcm3\times2\arcmin$), HST/WFC3 is an ideal instrument for detecting continuum information, including the slitless spectra of compact, passive galaxies in cluster cores \citep{Trump2011,Stanford2012}.

In this paper, we present the detection of a pair of massive quiescent galaxies at $z=2.244$/$z=2.242$ reside in the density peak of protocluster BOSS1244 with HST grism slitless spectroscopic observations.  Combing deep imaging and  NIR spectroscopy, we aim to understand how and when BCGs form and explore the galaxy properties in protocluster environments.  In Section~\ref{sec:observe}, we describe the observations and data reductions. In Section~\ref{sec:analyse}, we present the main analyses and results. We discuss our results in Section~\ref{sec:discuss}, and the conclusions are summarized in Section~\ref{sec:summary}.  Throughout this paper, we adopt the cosmological parameters of $\Omega_{\rm M} = 0.3$, $\Omega_{\rm \Lambda} =0.7$ and H$_0 = 70$\,km\,s$^{-1}$\,Mpc$^{-1}$ and magnitudes are presented in the AB system throughout this work.

\section{Observations and data reductions} \label{sec:observe}

The HST observations of five pointings in BOSS1244 were obtained between 17 January 2021 and 21 February 2021 by the MAMMOTH-Grism program (HST-GO-16276; P.I.: Xin~Wang). The FOV of each pointing  is $2\farcm3\times2\arcmin$. Each pointing contains three visits with different orientations. 
Each visit takes about one-orbit exposure, together with F125W pre-imaging.  The three grism exposures are combined to remove the blending of spectra in the BOSS1244 field. The total exposure times in F125W and G141 are 1,817\,s and 5,917\,s.  The data reduction is conducted using the {\it grizli} pipeline \citep{Brammer2021}. The detailed information about our WFC3/grism slitless observations and data reduction steps can be found in previous work \citep{Wang2021}.  In addition, we also use the archival WFC3 F160W imaging data (with an effective exposure time per pointing 2,614\,s) and the coverage region is the same as the F125W area. The image depths (5$\sigma$, point sources) in F125W and F160W are estimated from photometry within an aperture of radius 1$\arcsec$ on random positions in the blank background regions to be 24.95\,mag and 24.97\,mag, respectively.

\begin{figure}[!ht]
\centering
\includegraphics[trim=0mm 0mm 0mm 0mm,clip,height=0.4\textwidth]{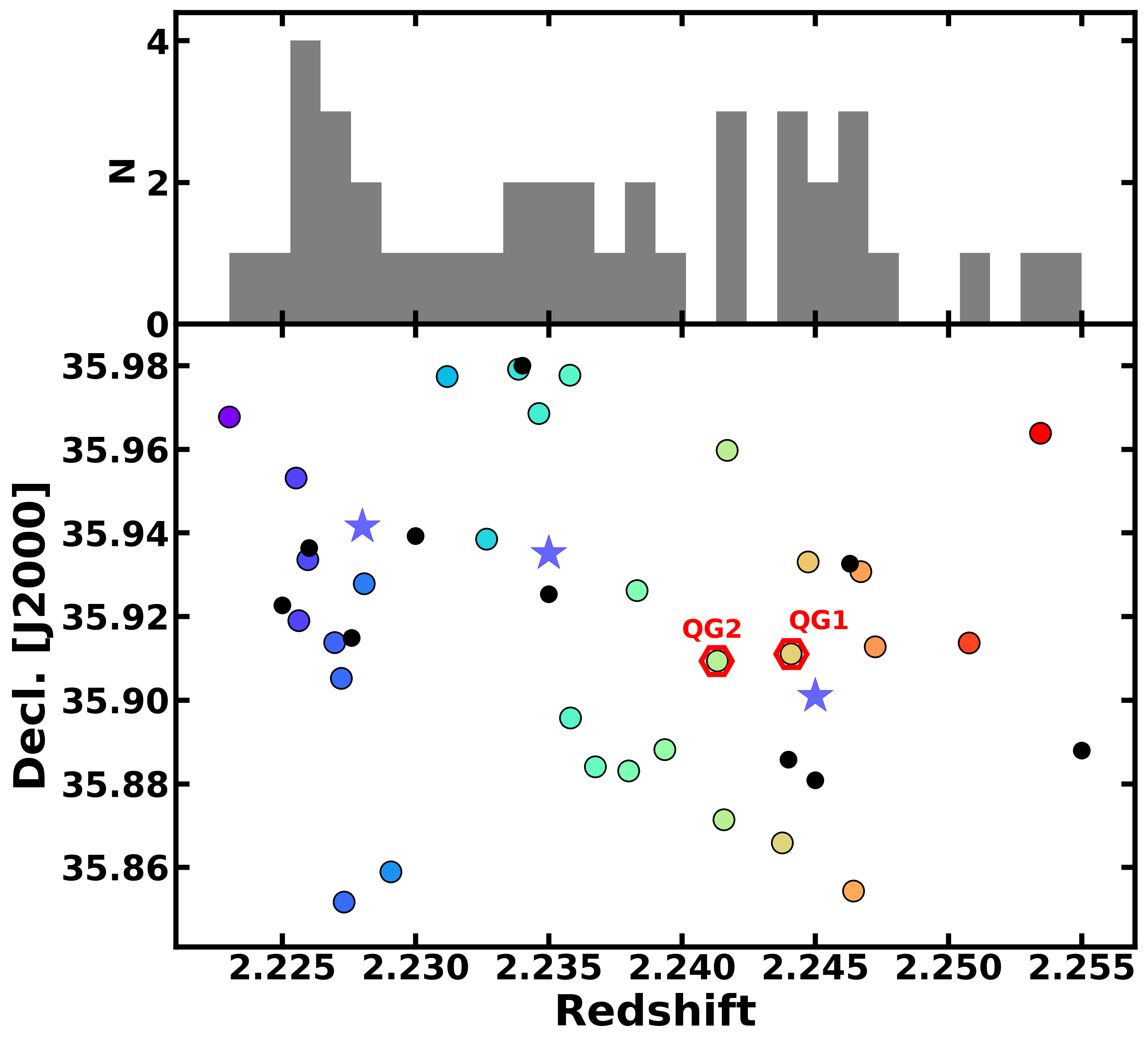}
\caption{Redshift distribution of 40 member galaxies against Declination in  the BOSS1244 SW region. The color-coded circles are  galaxies confirmed by grism at $z=2.223-2.255$. Black circles and blue stars represent the confirmed HAEs and quasars at $z=2.228-2.245$, respectively. Red hexagons denote two quiescent galaxies BOSS1244-QG1 and BOSS1244-QG2 labeled with ``QG1'' and ``QG2''. } 
\label{fig:fig2}
\end{figure}

Furthermore, our data sets in the BOSS1244 field include the deep ground-based imaging data taken with the Large Binocular Telescope (LBT)/LBC through SDT-$U_{\rm spec}$ and Sloan $z$-band in 2018A, Canada-France-Hawaii Telescope (CFHT)/WIRCam through narrowband H$_{2}$S(1) and broadband $K_{\rm s}$ in 2016A, and James Clerk Maxwell Telescope (JCMT)/SCUBA-2 at 850\,$\micron$ in 2018A and 2019A. The optical and NIR image data were reduced via the standard data reduction pipeline \citep{Schirmer2013, Zheng2021}, including the basic procedures of subtractions for bias and dark, flat-fielding, and removal of bad/hot pixels and sky subtraction. Source detection and photometry was performed using the software SExtractor \citep{Bertin1996}. SCAMP \citep{Bertin2006} was used to measure the astrometric calibration with bright stars, giving an accuracy of r.m.s.$\leq0\farcs1$ in these bands. We selected good-quality science images and stacked those of the same filter together with the help of the software tool SWarp \citep{Bertin2002}. The total exposure times in the SDT-$U_{\rm spec}$, $z$, H$_{2}$S(1) and  $K_{\rm s}$ bands are 4.65, 3.92, 7.50 and 7.50\,hours, and their 5$\sigma$ limiting magnitudes (within a 2$\arcsec$ diameter aperture) for  point sources are 26.67, 25.12, 22.58 and 23.29\,mag, respectively. The description of the JCMT/SCUBA-2 observations with integrated time of 18.67\,hours in BOSS1244 and data reduction were presented in \cite{Zhang2022}.

We extract one-dimensional (1D) grism slitless spectra in BOSS1244 and compute the redshifts of galaxies through the best fitting of spectral template synthesis. A total of 284 galaxies are identified with robust grism redshifts over $0.15<z<4$ \citep{Wang2021}. The redshift histogram shows a spike at $z=2.24\pm0.02$, which is consistent with ground-based NIR spectroscopy of HAEs \citep{Shi2021}. Here we choose galaxies at $z=2.223-2.255$ as the protocluster members, consistent with the redshift span detected by the NIR narrowband H$_{2}$S(1) filter used to identify the HAE candidates \citep{Zheng2021}. Combined with the previous ground-based NIR spectroscopy of HAEs, a total of 40 member galaxies and three QSOs are confirmed in the BOSS1244 southwest (SW) region. The spatial distribution of the 40 galaxies + 3 QSOs and the footprints of the five pointings of HST grism observations are shown in Figure~\ref{fig:fig1}, overlaid on the density maps of HAEs and SMGs in BOSS1244. Of the these objects, one QSO and one HAE are not covered by the HST observations. 
These members reside in volumes of $5481.30=8.92\times14.80\times41.52$\,cMpc$^{3}$. Namely, the scale of BOSS1244 SW region is 17.63\,cMpc, consistent with the typical size ($\sim15$\,cMpc) of a protocluster at $z\sim2.2$. Figure~\ref{fig:fig2} demonstrates the 40  member galaxies and three QSOs in the redshift versus Declination diagram.

\begin{figure*}[!ht]
\centering
\includegraphics[trim=0mm 0mm 0mm 0mm,clip,height=0.42\textwidth]{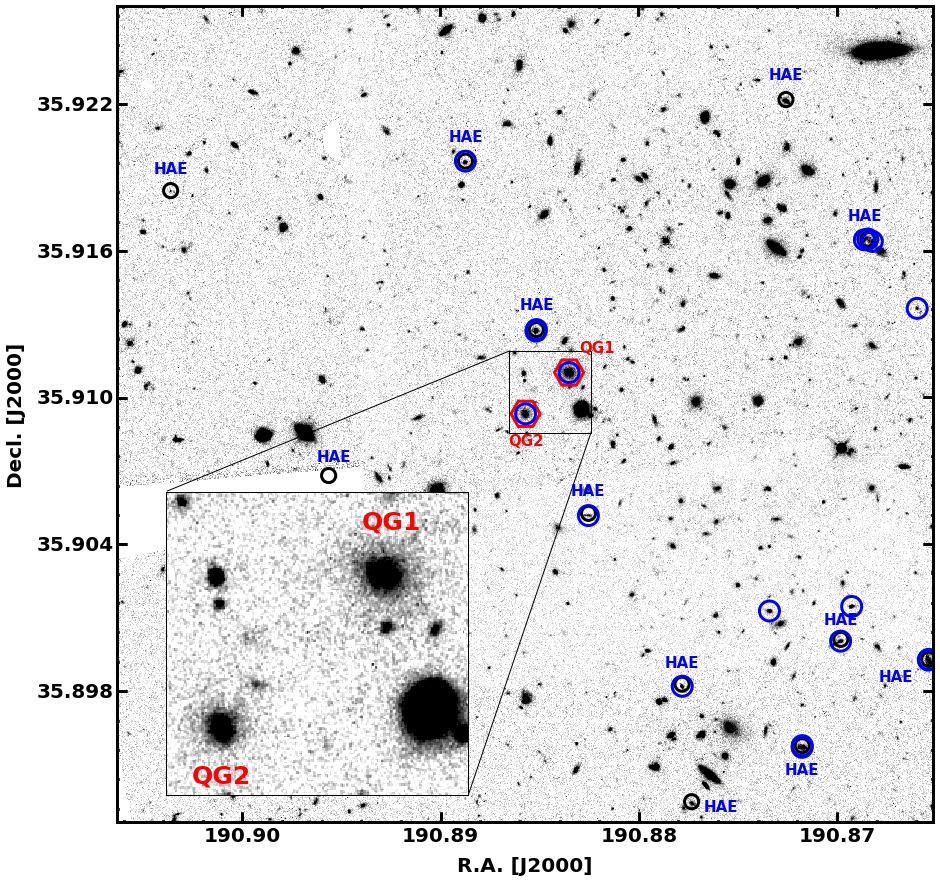}
\includegraphics[trim=0mm 0mm 0mm 0mm,clip,height=0.42\textwidth]{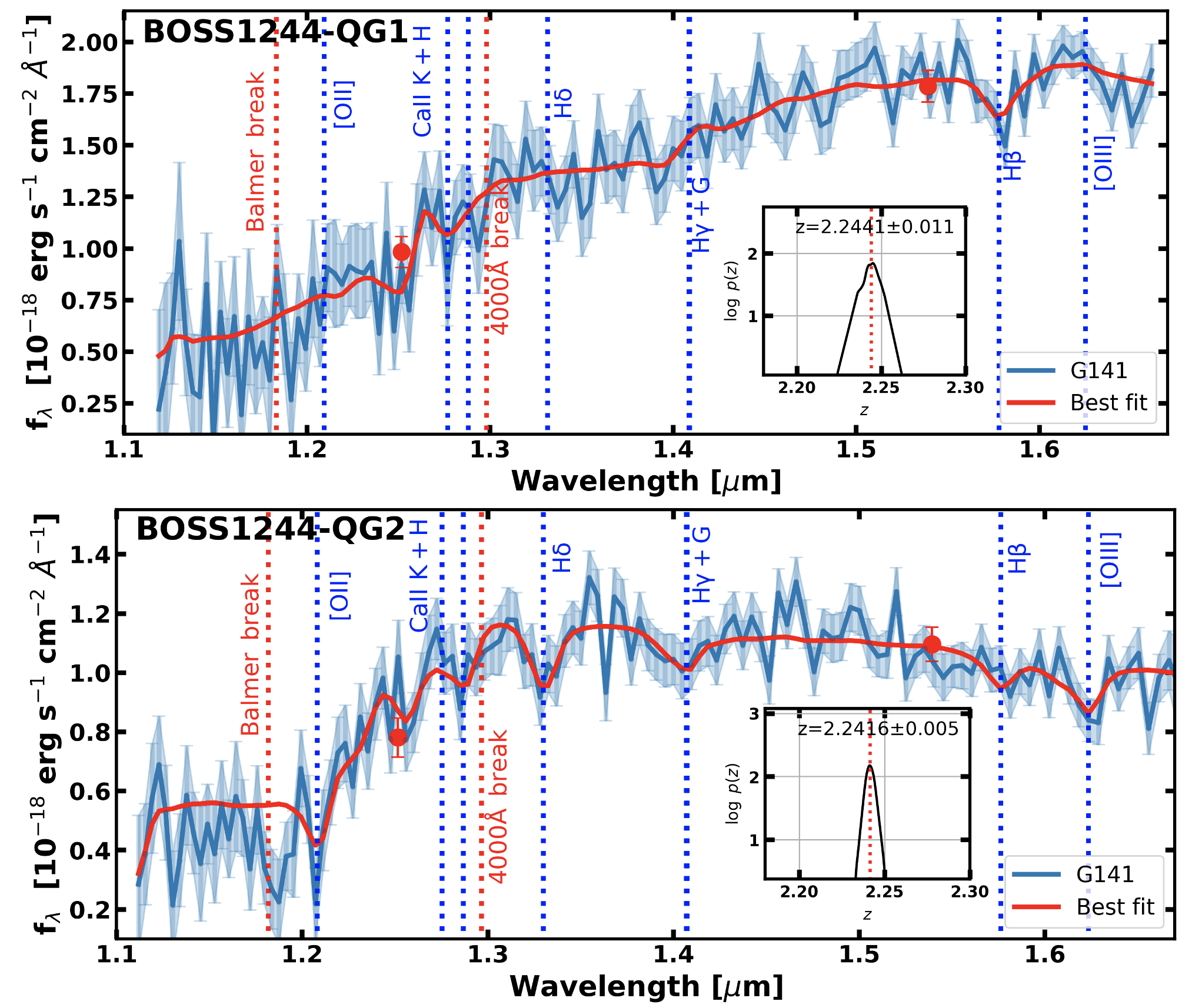}
\caption{ HST image (left) and one-dimensional grism spectroscopy (right) of BOSS1244-QG1 and BOSS1244-QG2.  The size of the left HST image is $2\farcm0 \times 2\farcm0$. The inset box ($12\arcsec \times 12\arcsec$) is a zoom-in on the BOSS1244-QG1 and BOSS1244-QG2.  Black circles mark the HAE candidates, blue circles denote the spectroscopically confirmed galaxies at $z=2.24$.  Over this region the  overdensity factor of HAEs is $\delta_{g}>40$. The red hexagons mark QG1 and QG2. In the right panels, light blue curves with 1$\sigma$ uncertainties refer to the extracted spectra from HST/WFC3 grism observations acquired by the MAMMOTH-Grism survey. Red curves are the best fitting with lines based on the Flexible Stellar Population Synthesis (FSPS) template models. The  expected spectral lines (e.g., Balmer or Ca\,{\small II} absorption) are marked with the vertical blue dotted lines. Red dotted lines show the location of Balmer and 4000\,\AA\ break. The inner panel presents the redshift probability distribution function for BOSS1244-QG1/BOSS1244-QG2. } 
\label{fig:fig3}
\end{figure*}

Among the 40 spectroscopically-identified galaxies, two are not HAEs. They are identified as quiescent galaxies and reside in the density peak region of the BOSS1244 SW protocluster. The two targets are named as BOSS1244-QG1 and BOSS1244-QG2, and labeled with ``QG1'' and ``QG2'' in Figure~\ref{fig:fig1} and \ref{fig:fig2}. The HST images and grism slitless spectra of the two galaxies are shown Figure~\ref{fig:fig3}.  BOSS1244-QG2 exhibits clear Balmer (H$\delta$, H$\gamma$, H$\beta$) and Ca\,{\small II} absorption, while BOSS1244-QG1 shows strong H$\beta$ absorption line and other weak Balmer and Ca\,{\small II} absorption features. No emission lines are detected. BOSS1244-QG1 and BOSS1244-QG2 were optimally extracted and simultaneously fitted with a suite of galaxy templates. We fitted the Flexible Stellar Population Sysnthesis (FSPS) template models with redshift over a fine ($\Delta z = 0.0004$) grid from $z = 0.2-4.0$, giving the probability distribution function of redshifts. Following \cite{Willis2020}, we defined a robust redshift using $P(z)>0.5$ where $P(z)$ was the integral of redshift probability distribution function for BOSS1244-QG1/BOSS1244-QG2 over the interval $2.223<z<2.267$. The estimated redshifts of BOSS1244-QG1 and BOSS1244-QG2 are $z=2.2441\pm 0.011$ and $z=2.2416\pm 0.005$, respectively. They belong to protocluster BOSS1244. A detailed analysis of the physical  properties of these two galaxies are presented in Section~\ref{sec:qg12}.

\section{ Analysis and Results} \label{sec:analyse}
 
\subsection{Structural parameters of member galaxies } 

We measure structural parameters from the F160W images for 39 member galaxies in BOSS1244, including our two target quiescent galaxies BOSS1244-QG1 and BOSS1244-QG2. Note that one HAE in our sample is not covered by the HST imaging observations. We use SExtractor to detect these galaxies and extract their photometric and astrometric parameters, including coordinates, magnitudes, effective radius, axis ratio and position angle as the initial input parameters for GALFIT \citep{Peng2002, Peng2010}. We mask the objects close to the targets taking advantage of a segmentation map from SExtractor. GALFIT is run iteratively to obtain the final structural parameters  of the measured galaxies.

\begin{figure}[!ht]
\centering
\includegraphics[width=\columnwidth]{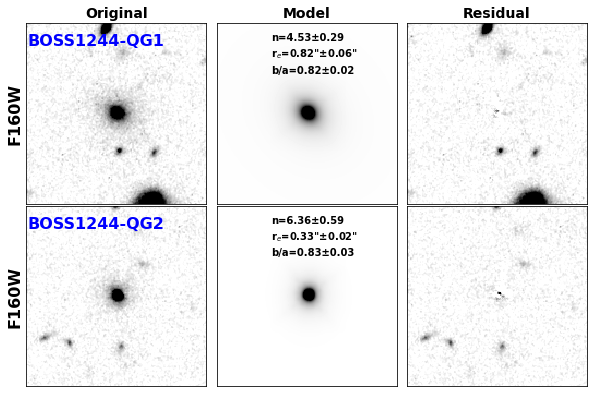}
\caption{GALFIT models to BOSS1244-QG1 and BOSS1244-QG2 in F160W. The three columns from left to right are the original image, the GALFIT model, and the residual image. S{\'e}rsic index $n$, effective radius $r_{\rm e}$, and axis ratio are marked in model image. The size of each postage stamp image is $10\arcsec \times 10\arcsec$.} 
\label{fig:figgalfit}
\end{figure}

BOSS1244-QG1 is the brightest galaxy in our sample of 40 member galaxies and BOSS1244-QG2 is the second brightest.  We focus on BOSS1244-QG1 and BOSS1244-QG2 and list their best-fitting structural parameters and total magnitudes in Table~\ref{tab:taba}.  Figure~\ref{fig:figgalfit} presents the HST F160W images, best-fitting  S{\'e}rsic models and residual images of the two galaxies. One can see that both of the two galaxies can be well fitted by a single S{\'e}rsic profile, giving the best-fitting S{\'e}rsic index $n=4.53\pm0.29$  ($n=6.36\pm0.59$) and half-light radius $R_{\rm e}=6.76\pm0.50$\,kpc ($R_{\rm e}=2.72\pm0.16$\,kpc) for  BOSS1244-QG1 (BOSS1244-QG2).  Clearly the two quiescent galaxies at $z\sim 2.24$ are similar to early-type galaxies in the local Universe in terms of their structural parameters.

\begin{table}
\centering
\caption{The physical properties of BOSS1244-QG1 and BOSS1244-QG2.} \label{tab:taba}
\begin{tabular}{lrr}
\hline 
\hline
 Properties & BOSS1244-QG1 & BOSS1244-QG2  \\
 \hline
R.A.\,(J2000.0)   &  12:43:32.04  & 12:43:32.57 \\ 
Decl.\,(J2000.0)  & +35:54:39.68  & +35:54:33.59 \\
spec-$z$  &   2.244$\pm$0.011 & 2.242$\pm$0.005  \\
K$_{\rm s}\,$(mag) & 20.59$\pm$0.02 & 21.17$\pm$0.02 \\
H$_{2}$S(1)\,(mag) & 20.58$\pm$0.09 & 21.12$\pm$0.05 \\
F125W\,(mag) &  22.13$\pm$0.07 & 22.38$\pm$0.03  \\
F160W\,(mag) &  21.03$\pm$0.04 & 21.56$\pm$0.03  \\
r$_{\rm e,F160W}$\,($\arcsec$) & 0.82$\pm$0.06  & 0.33$\pm$0.02 \\
r$_{\rm e,F160W}$\,($\rm kpc$)  & 6.76$\pm$0.50 & 2.72$\pm$0.16 \\
$n$ (F160W)  & 4.53$\pm$0.29  & 6.36$\pm$0.59\\
$b/a$ (F160W)   & 0.82$\pm$0.02 & 0.83$\pm$0.03 \\
P.A. (F160W)   & 42.02$\pm$3.53 & 0.53$\pm$5.17 \\
D$_{\rm n}$4000 index   & 1.47$\pm$0.29  &  1.24$\pm$0.15 \\
log(M$_{*}$/M$_{\odot}$)  & $11.62_{-0.07}^{+0.08}$ & $10.85_{-0.03}^{+0.13}$ \\
log(SFR/$M_{\odot}$\,yr$^{-1}$)  & $0.69_{-0.38}^{+0.33}$ & $-0.51_{-0.11}^{+0.76}$ \\
Age (Gyr)  & $2.14_{-0.19}^{+0.15}$ & $0.87_{-0.27}^{+0.02}$ \\
$A_{V}$\,(mag)  & $0.67_{-0.30}^{+0.26}$ & $0.01_{-0.01}^{+0.62}$ \\
log($\left\langle \rm SFR_{\rm main} \right\rangle$/$M_{\odot}$\,yr$^{-1}$)$^{\rm a}$ & 2.94$_{-0.09}^{+0.05}$ & 2.64$_{-0.02}^{+0.22}$ \\
log($t_{50}$/yr)$^{\rm b}$ & 9.23$_{-0.05}^{+0.14}$ & 8.87$_{-0.17}^{+0.01}$ \\
log($t_{\rm q}$/yr)$^{\rm c}$ & 8.93$_{-0.11}^{+0.17}$ & 8.67$_{-0.22}^{+0.01}$ \\
log($\tau$/yr)  & $8.45_{-0.04}^{+0.05}$ &  $7.94_{-0.06}^{+0.05}$ \\
\hline
\end{tabular}
\tablecomments{ The physical properties of BOSS1244-QG1 and BOSS1244-QG2 are derived from SED fitting with FAST++. $^{\rm a} \left\langle \rm SFR_{\rm main} \right\rangle$ is the average SFR during the shortest time interval over which 68\% of SFR took place, $^{\rm b} t_{\rm 50}$ is the elapsed time since the galaxy had formed 50\% of the total stellar mass, and $^{\rm c} t_{\rm q}$ is the elapsed time since SFR drops below 10\%  of $\left\langle \rm SFR_{\rm main} \right\rangle$.}
\end{table}

\begin{figure*}[!ht]
\centering
\includegraphics[trim=0mm 0mm 0mm 1mm,clip,height=0.25\textwidth]{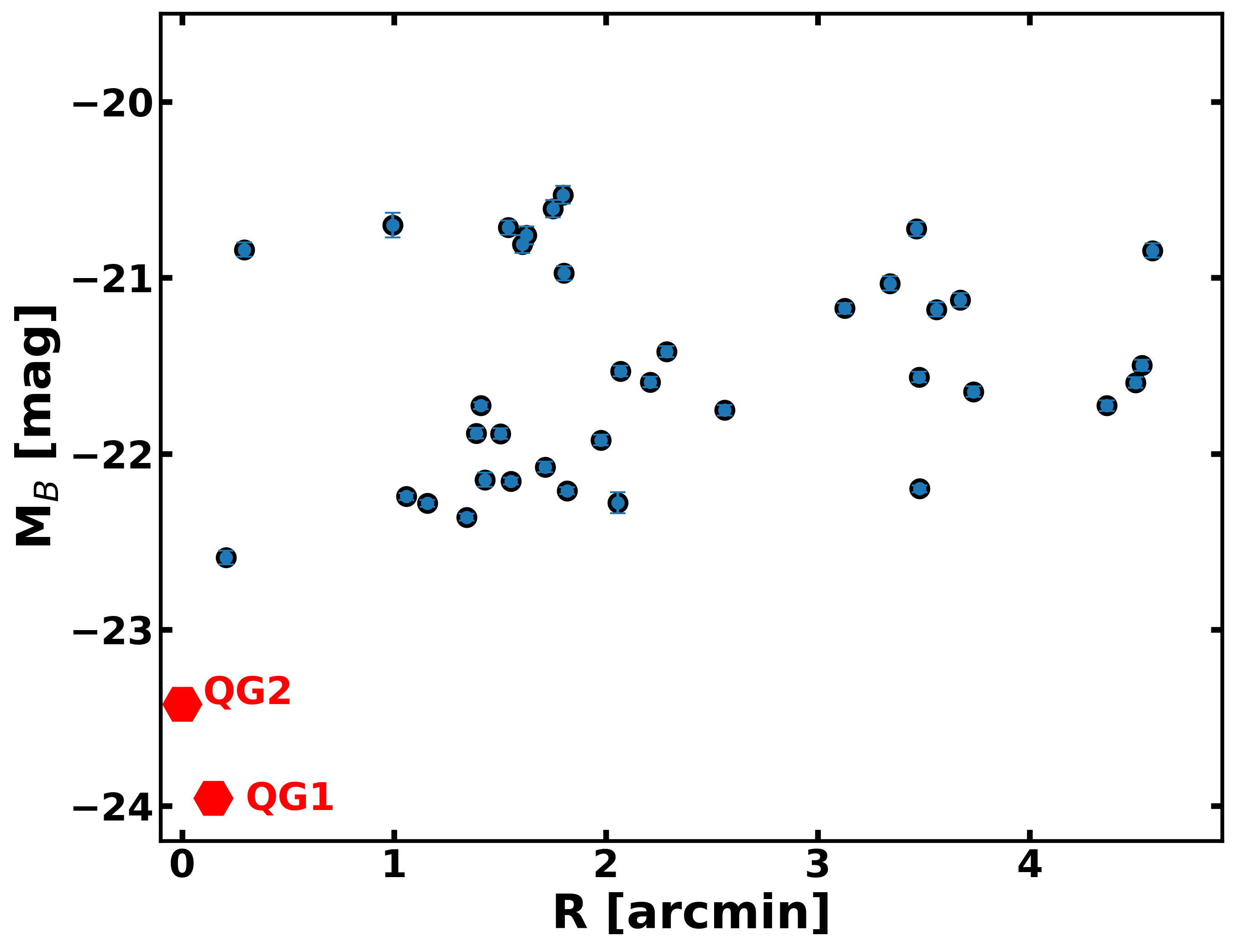}
\includegraphics[trim=0mm 0mm 0mm 1mm,clip,height=0.25\textwidth]{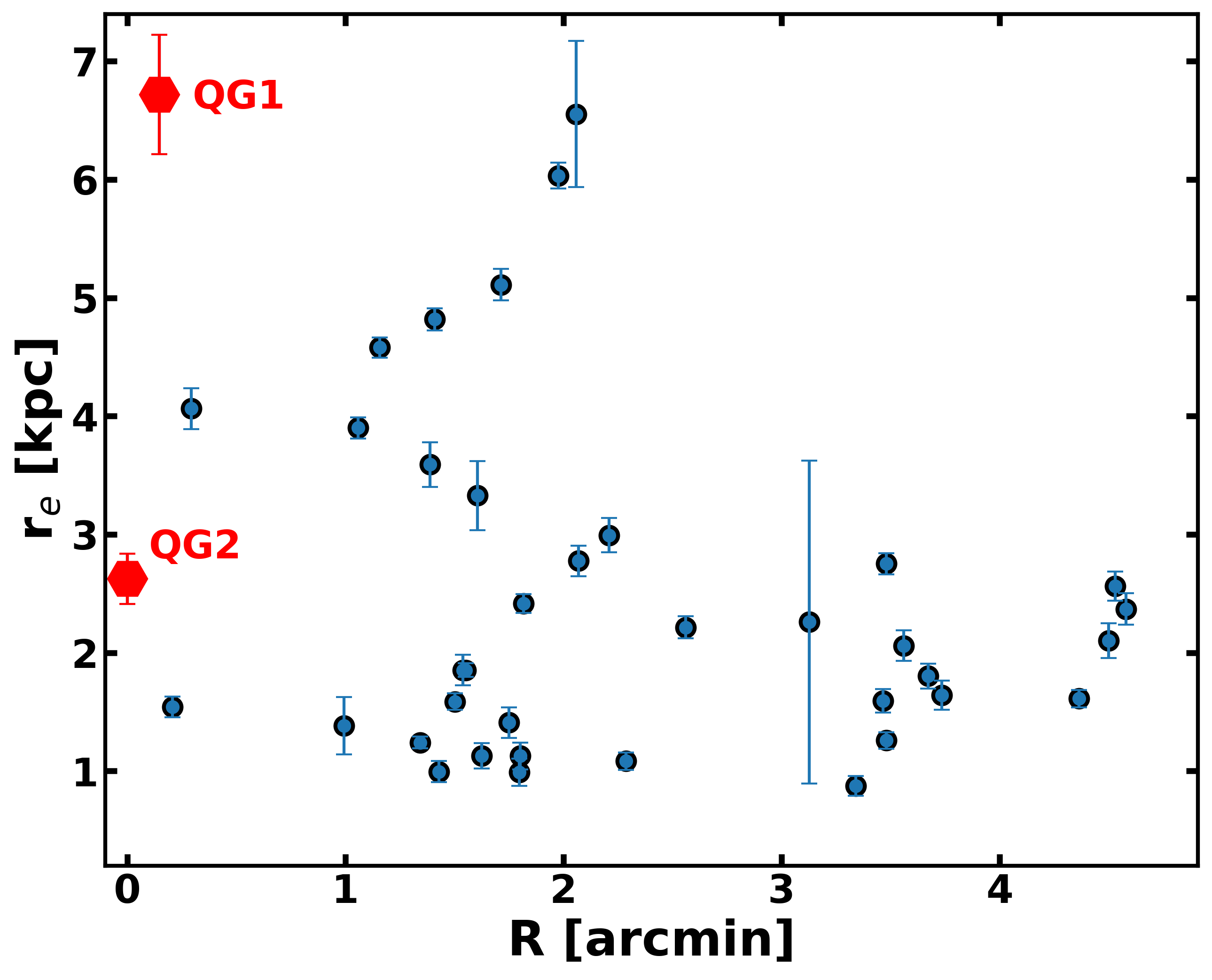}
\includegraphics[trim=0mm 0mm 0mm 1mm,clip,height=0.25\textwidth]{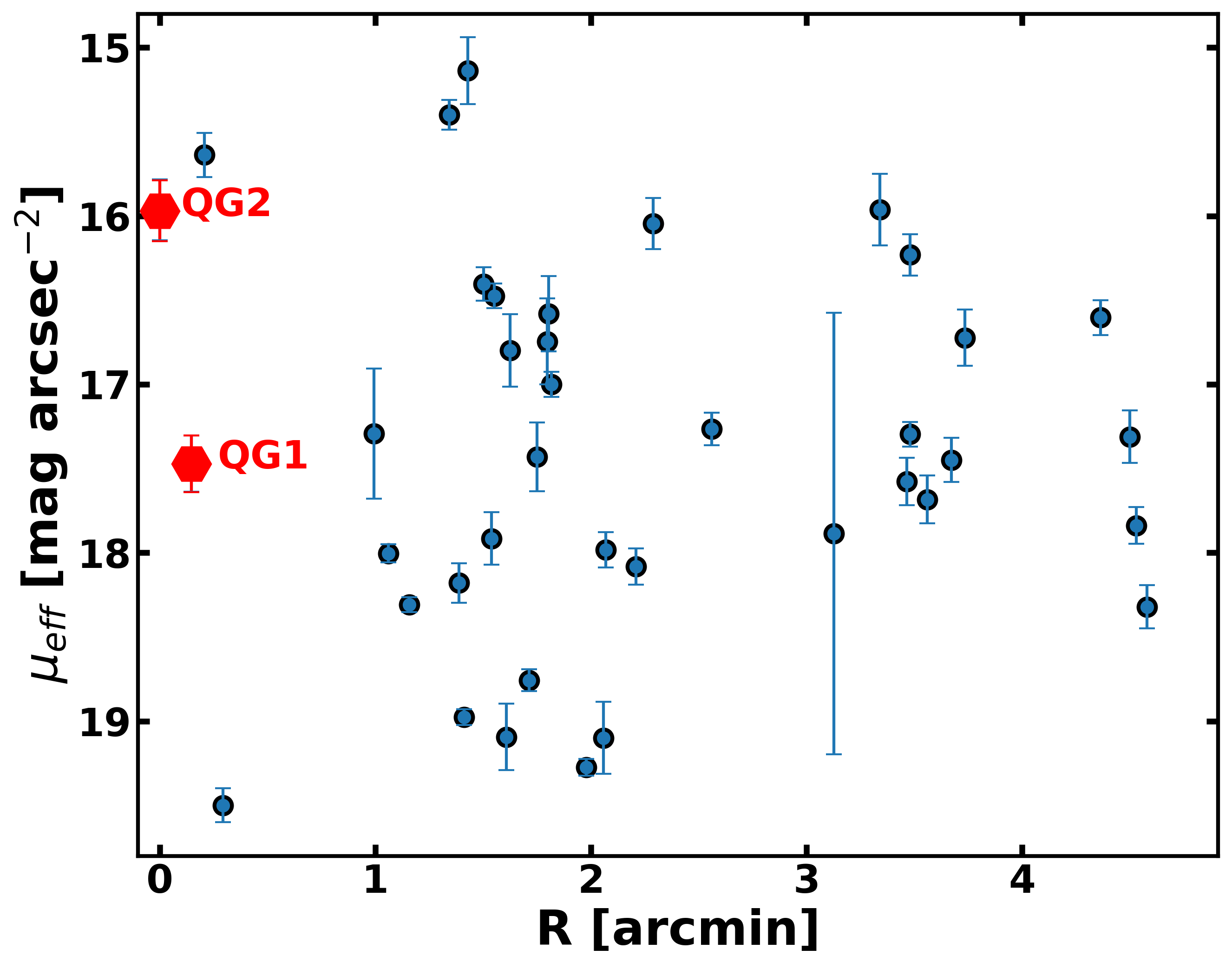}
\caption{Distribution of the rest-frame $B$-band absolute magnitude (left), effective radius $r_{\rm e}$ (middle), and  effective surface brightness $\mu_{\rm eff}$ (right) for 39 spectroscopically-confirmed galaxies with $z=2.223-2.255$  located at different radial distances in BOSS1244 SW region. Note that all extended galaxies with $r_{\rm e}>3$\,kpc are located in the central area of $R<\sim 2\arcmin$. }
\label{fig:fig4}
\end{figure*}

\begin{figure}[!ht]
\centering
\includegraphics[width=\columnwidth]{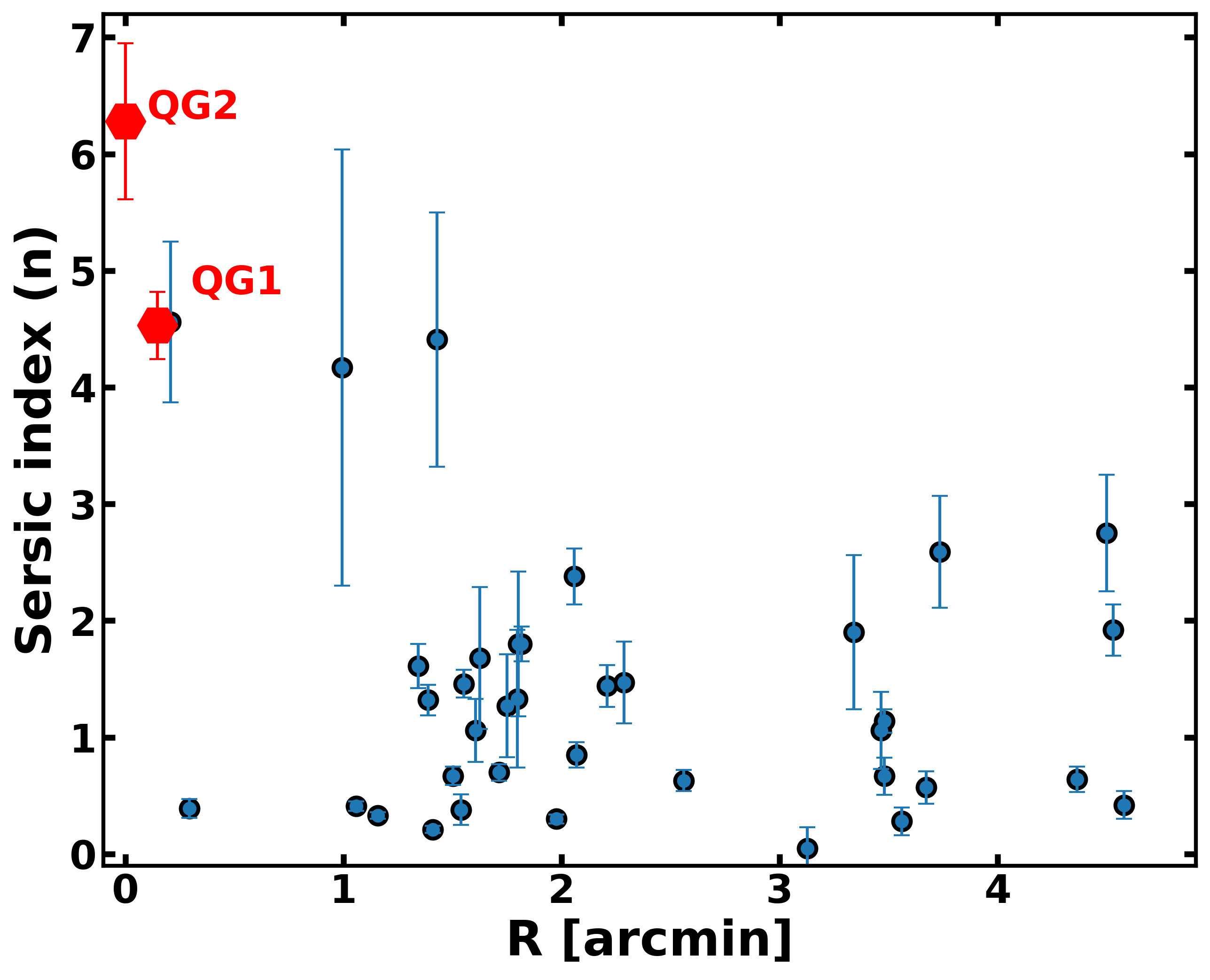}
\caption{S{\'e}rsic index $n$ as a function of radial distance for 39 spectroscopically-confirmed galaxies with $z=2.223-2.255$ in the BOSS1244 SW region.  Clearly, five bulge-dominated galaxies ($n>3$) are all located in the central area of $R<2\arcmin$.  } 
\label{fig:fig4b}
\end{figure}

As can be seen from Figure~\ref{fig:fig1}, BOSS1244-QG1 and BOSS1244-QG2 are  located roughly at the center of the SW region. We intend to examine if the properties of these confirmed member galaxies show dependence  on density.  The density contours presented in Figure~\ref{fig:fig1} are used to derive the effective radius of a circle with area equal to that of the region enclosed by the density contour at the galaxy. Doing so,  we obtain the effective radial distance as an indicator  of density ranging from the center to  the outskirts of the BOSS1244 SW region.  The position of BOSS1244-QG2 is chosen to be the center. We point out that our results are not sensitive to the selection of the center position.

Figure~\ref{fig:fig4} shows the distribution of 39 spectroscopically-confirmed galaxies with $z=2.223-2.255$ in the rest-frame $B$-band absolute magnitude M$_{B}$,  half-light radius $r_{\rm e}$ and the effective surface brightness $\mu_{\rm eff}$ in comparison with radial distance. We obtained M$_{B}$ and $\mu_{\rm eff}$ using the equations of  $M_{\rm B}=m-5$\rm log($d$[\rm Mpc])$-25.0+2.5$log(1+$z$) and  $\mu_{\rm eff}=m+ 2.5\rm log(2\pi r_{\rm e}^{2}$)$-10$log(1+$z$), where m is the total apparent magnitude, $d$ is the luminosity distance and $-10$log(1+$z$) is the cosmological dimming correction term. One can see that the majority of the sample galaxies have $-22.5<M_{B}<-20.5$, but BOSS1244-QG1 and BOSS1244-QG2 are roughly two magnitudes brighter than the typical brightness, making the two quiescent galaxies distinct from the rest. There are 11 galaxies with $r_{\rm e}>3$\,kpc, being already sufficiently large in size in comparison with Milky Way-like galaxies ($r_{\rm e}\sim 3$\,kpc) in the local Universe.  BOSS1244-QG1 appears as the largest one among the member galaxies. Note that the largest galaxies are all residing in the central densest region of $R<2\arcmin$. On the other hand,  the observed effective surface brightness (without correction for dust attenuation) of these galaxies mostly distributes in a range of  $\mu_{\rm eff}\sim 15-20$\,mag\,arcsec$^{-2}$,  being consistent with that of typical star-forming galaxies at $z\sim 2-3$ \citep{vanderWel2014}.

We demonstrate the dependence of S{\'e}rsic index $n$ on the radial distance for our sample of 40 galaxies in Figure~\ref{fig:fig4b}.  
Bulge-dominated galaxies make up 20\% of the sample (including BOSS1244-QG1 and BOSS1244-QG2).
Among them, about $16.2\pm6.6$ percent of member SFGs belong to the bulge-dominated ones. The latter is consistent with the case  in the virialized cluster XLSSC122 at $z\sim2$ \citep{Noordeh2021}.  
Strikingly, all five member galaxies with $n>3$ are  located within the central region of $R<2\arcmin$.  This radius corresponds  to a physical size of $\sim1$\,Mpc at $z=2.24$.  
From Figure~\ref{fig:fig4} and Figure~\ref{fig:fig4b}, one can see a gradual transition from the center to the outskirts of the BOSS1244 SW structure: the presence of two massive quiescent galaxies at the center, concentration of large-size and bulge-dominated galaxies within  $R<\sim2\arcmin$, and extended star-forming galaxies at $R>2\arcmin$. One can easily associate the formation of the galaxies with $n>3$ and/or $r_{\rm e}>3$\,kpc with the assembly of the core of the BOSS1244 SW structure. 

\begin{figure}[!ht]
\centering
\includegraphics[trim=0mm 0mm 0mm 0mm,clip,width=\columnwidth]{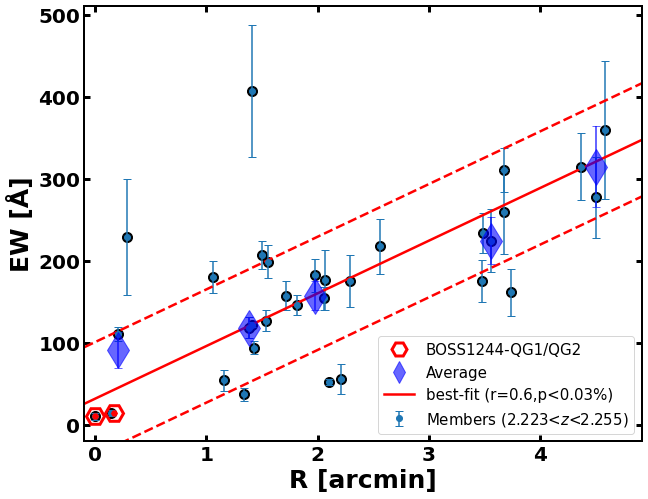}
\caption{ H$\alpha$ equivalent width (EW) versus radial distance relative to the quiescent galaxy BOSS1244-QG2 for member galaxies.  The light blue data points are the spectroscopically confirmed galaxies at $z=2.223-2.255$ in the BOSS1244 SW region and the blue diamonds are the average values in different bins. The red hexagons are BOSS1244-QG1 and BOSS1244-QG2. The red solid line represents the robust fit to the whole of member galaxies with errors. The red dashed lines show the 3$\sigma$ scatter. A strong correlation exists between EW and radial distance, suggesting that the gradual suppression star formation in galaxies coincides with the increase of environment density. } 
\label{fig:fig5}
\end{figure}

\subsection{A prominent star formation gradient} 

We also aim to examine the dependence of galaxy star formation on the radial distance in BOSS1244 SW structure.  
The equivalent width (EW) of  H$\alpha$ is a good indicator of galaxy star formation rate (SFR) relative to its stellar mass, i.e., specific SFR (sSFR), so we measure the H$\alpha$ EW through NIR narrowband H$_{2}$S(1) and broadband $K_{\rm s}$. The rest-frame EWs of member galaxies  in BOSS1244 with the following equations.

\begin{equation}
\centering
f_{\rm c}=\frac{f_{K_{\rm s}}-f_{\rm H_{2}S(1)}(\Delta \lambda_{\rm H_{2}S(1)}/\Delta \lambda_{K_{\rm s}})}{1-\Delta \lambda_{\rm H_{2}S(1)}/\Delta \lambda_{K_{\rm s}}},
\end{equation}

\begin{equation}
\centering
EW=(1+z)^{-1}\frac{F_{\rm line}}{f_{\rm c}}.
\end{equation}
where $\Delta \lambda_{\rm H_{2}S(1)}=0.0295\,\mu$m and $\Delta\lambda_{K_{\rm s}}=0.327\,\mu$m are the the full width at half-maximum (FWHM) of the H$_{2}$S(1) and $K_{\rm s}$ filters, and $f_{\rm H_2S(1)}$ and $f_{K_{\rm s}}$ are the flux density in the H$_{2}$S(1) and $K_{\rm s}$-band, respectively.  Note that 32 of 40 member galaxies are detected in the H$_{2}$S(1) and $K_{\rm s}$ bands. For BOSS1244-QG1 and BOSS1244-QG2, their H$\alpha$ EWs are 2.60$\pm$10.26\,\AA \,\,and 6.48$\pm$6.20\,\AA, respectively, suggesting that BOSS1244-QG1 and  BOSS1244-QG2 indeed have no or very weak star formation.

Importantly, we find a prominent gradient in galaxy star formation intensity from the center to the outskirts of BOSS1244.  Figure~\ref{fig:fig5} shows  the H$\alpha$ EW of  member galaxies at increasing radial distance from the location of BOSS1244-QG2. One can see that sSFR decreases at decreasing radial distance from the outskirts to the center (i.e., increasing density). The two quiescent galaxies  BOSS1244-QG1 and BOSS1244-QG2 have ceased star formation at the center. The relationship formed by these data points, characterized by a Spearman/Pearson correlation coefficient $r=0.60\pm0.05/0.61\pm0.07$ and  $P-$value of $0.00033/0.00019$, turns to be very significant correlation.  The relation is given over a scale of 17.63\,cMpc, consistent with the typical scale ($\sim15\,$cMpc) of massive protoclusters at $z\sim2-3$. The steep decrease of sSFR with increasing density reveals that the suppression degree of star formation in galaxies becomes stronger towards the core region of the BOSS1244 SW structure. In other words,  galaxies in the BOSS1244 SW region was undergoing star formation quenching from the center gradually to the outskirts.  This strong star formation gradient is likely the most prominent observed evidence for star formation quenching coupled with the formation of a mature cluster core in protoclusters at $z>2$.

\subsection{The physical properties of BOSS1244-QG1 and BOSS1244-QG2} \label{sec:qg12}

We constrain the physical properties of BOSS1244-QG1 and BOSS1244-QG2 using the multiwavelength photometry and spectroscopy.  The properties of other member galaxies are presented in the future work (Shi,D.D et al. in preparation). Considering BOSS1244-QG1 and BOSS1244-QG2 are relatively compact and isolated objects, their total intrinsic fluxes in multiwavelength data are measured through deconvolution point spread function (PSF) models. We first select several bright, isolated stars and construct a normalized PSF in each band, then the total magnitude and other structural parameters are derived from the best-fitting S{\'e}rsic profile to the F160W image through GALFIT \citep{Peng2010}, and finally obtain the photometric fluxes of these two QGs in other bands by setting the same parameters as for the F160W band. For signals that are too faint to be securely detected in a certain band (e.g., SDT-$U_{\rm spec}$), we use forced photometry ($2-3$ times FWHM of seeing aperture is fixed) to estimate the upper limits of their fluxes.

The 4000\AA\,break is produced by a combination of metal absorption on the atmosphere of old and cool stars and the lack of flux from young and hot OB stars. The strength of the 4000\AA\,break is used to trace the age of galaxies. 
We use D$_{\rm n}$4000 index to quantify the strength of the 4000\AA\,break; D$_{\rm n}$4000 index is defined as follows \citep{Balogh1999}:
\begin{equation}
\centering
D_{n}4000=\sum_{\small\lambda=4000\,\rm\AA}^{\small 4100\,\rm\AA}F_{\lambda}/\sum_{\small\lambda=3850\,\rm\AA}^{\small 3950\,\rm\AA}F_{\lambda}, 
\end{equation}
where $F_{\lambda}$ is the flux per unit wavelength.

The measured D$_{\rm n}$4000 index in BOSS1244-QG1 is $1.47\pm0.29$, which is slightly larger than BOSS1244-QG2 with $1.24\pm0.15$, suggesting that BOSS1244-QG1 is older than BOSS1244-QG2. This is consistent with the redder color of BOSS1244-QG1 (F125W-F160W=1.10$\pm$0.08) compared to BOSS1244-QG2 (F125W-F160W=0.82$\pm$0.04). 

We constrain the star formation histories (SFHs) of BOSS1244-QG1 and BOSS1244-QG2 using spectral energy distribution (SED) fitting techniques. We use the software Fitting and Assessment of Synthetic Templates (FAST++)\footnote{\url{https://github.com/cschreib/fastpp}}  to obtain the stellar properties of BOSS1244-QG1 and BOSS1244-QG2. This is a full rewriten FAST code (FAST++ v1.3) that can handle much larger parameter grids and generate models with arbitrary SFHs.  We adopt \cite{Bruzual2003} stellar population model, the \cite{Chabrier2003} IMF,  the \cite{Calzetti2000} dust attenuation law and Solar metallicity ($Z=0.02$) to fit the SEDs of BOSS1244-QG1 and BOSS1244-QG2. We set extinction $A_{\rm V}$ ranging over $0-4$\,mag with a step of 0.01\,mag and stellar age ranging over $6.0-10.3$\,dex  in units of year with a step of 0.1\,dex. The delayed exponentially declining SFH (SFR$\sim t \times \rm exp(-t/\tau)$) is used, where $t$ is time from the formation. The broadband SED and spectrum can be simultaneously fitted with FAST++. We run FAST++ with a velocity dispersion between 300 and 1000\,km\,s$^{-1}$ with steps of 50\,km\,s$^{-1}$. Note that the redshift is fixed at $z=2.2441/2.2416$ derived from the grism spectra.

\begin{figure*}[!ht]
\setlength{\abovecaptionskip}{0pt}
\begin{center}
\includegraphics[trim=0mm 0mm 0mm 1mm,clip,height=0.45\textwidth]{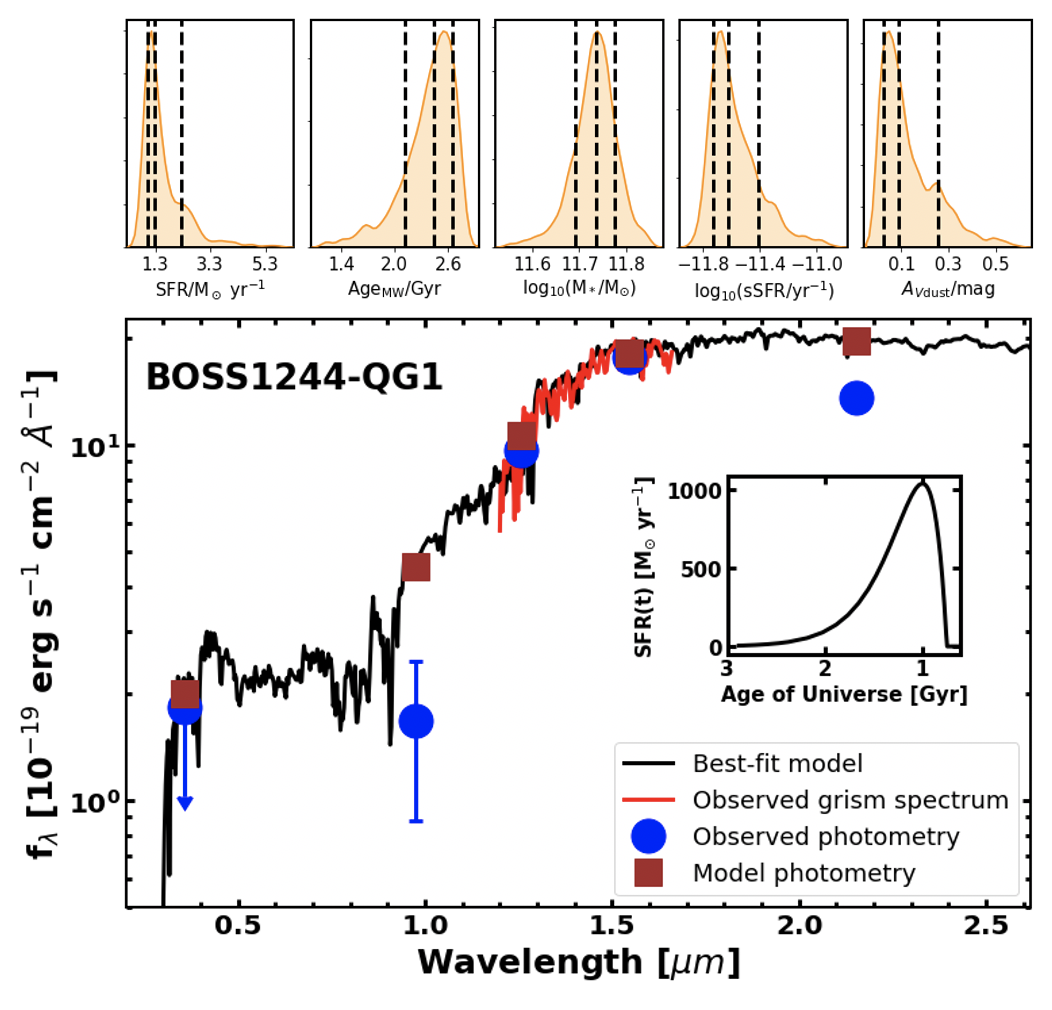}
\includegraphics[trim=0mm 0mm 0mm 1mm,clip,height=0.45\textwidth]{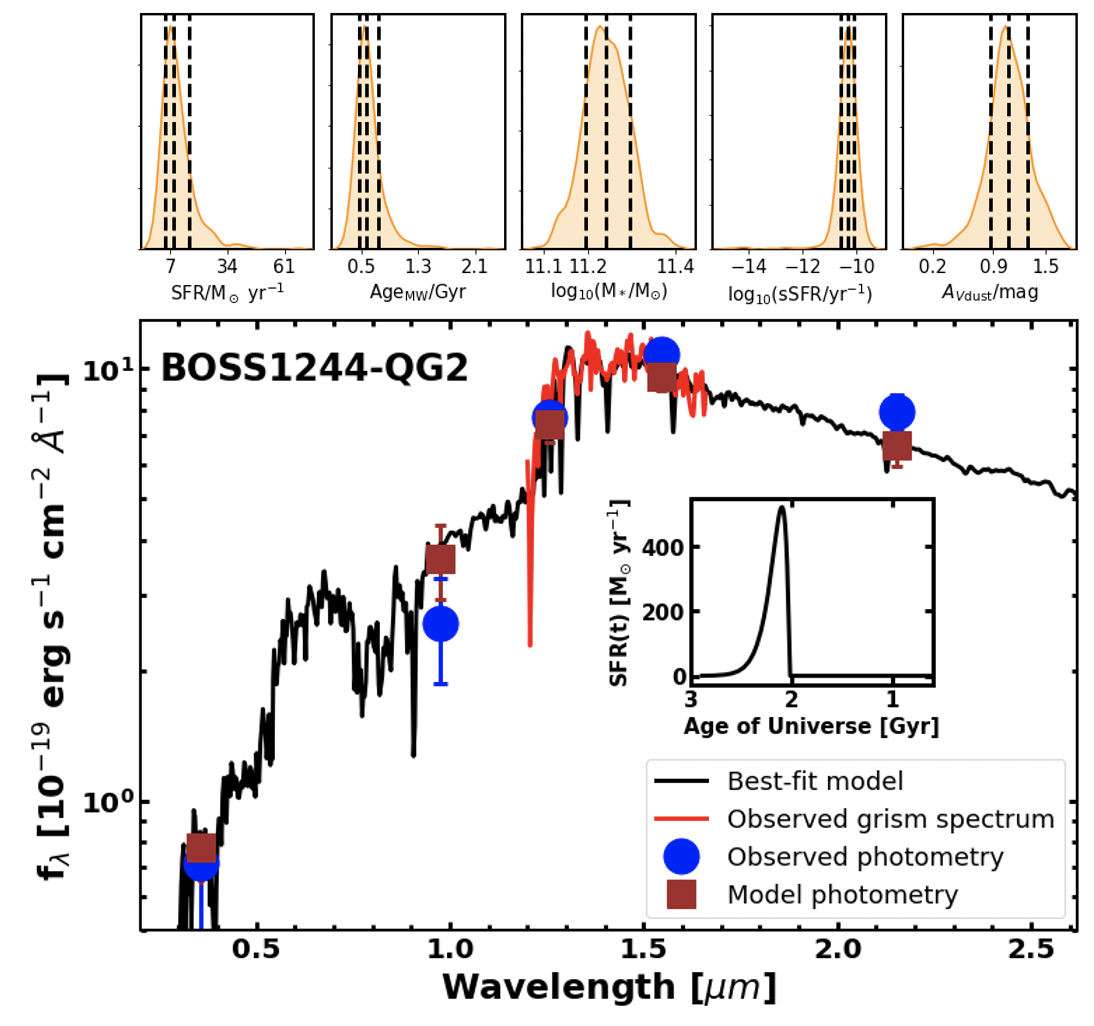}
\caption{SED fitting of BOSS1244-QG1 (left) and BOSS1244-QG2 (right). The brown squares are the model photometry and the blue points are the observed photometry for BOSS1244-QG1 and BOSS1244-QG2. The observed HST grism spectrum is shown in red line, and the black curve in left/right represents the best-fit model from FAST++. The inner panel  in left/right gives the SFH of BOSS1244-QG1/BOSS1244-QG2. The upper panels in left/right show the probability distributions of the SFR, mass-weighted age, $M_{\ast}$, sSFR and $A_{V}$ from the SED fitting results of BAGPIPES. The 16th, 50th and 84th percentiles marked by the black dashed lines. } 
\label{fig:fige2}
\end{center}
\end{figure*}

The best-fit models of BOSS1244-QG1 and BOSS1244-QG2 with FAST++  are shown in Figure~\ref{fig:fige2}, and the reduced $\chi^2$ is 1.30 and 1.46, respectively.  For BOSS1244-QG1, we obtain stellar mass $M_{\ast}=4.19_{-0.62}^{+0.84}\times10^{11}$\,M$_{\odot}$, stellar age $t=2.14_{-0.19}^{+0.15}$\,Gyr, dust attenuation $A_{V}=0.67_{-0.30}^{+0.26}$\,mag and SFR$=4.90_{-2.86}^{+5.57}$\,$M_{\odot}$\,yr$^{-1}$. For the BOSS1244-QG2, we obtain $M_{\ast}=7.08_{-0.47}^{+2.47}\times10^{10}$\,M$_{\odot}$, $t=0.87_{-0.27}^{+0.02}$\,Gyr, $A_{V}=0.01_{-0.01}^{+0.62}$\,mag and SFR$=0.31_{-0.07}^{+1.47}$\,$M_{\odot}$\,yr$^{-1}$. The other physical properties of BOSS1244-QG1 and BOSS1244-QG2 are also listed in Table~\ref{tab:taba}. The uncertainties of the values from the SED resullts in Table~\ref{tab:taba} are the 95\% confidence interval ($2\sigma$) values. In addition, we find that when the metallicity ($Z$) is set as a free parameter, the metallicities ($Z$) of BOSS1244-QG1 and BOSS1244-QG2 are in the range of $0.02-0.04$. This indicates that BOSS1244-QG1 and BOSS1244-QG2 are massive and metal-rich galaxies. 

We also use the software tool Bayesian Analysis of Galaxies for Physical Inference and Parameter EStimation (BAGPIPES) to double-check the stellar properties of BOSS1244-QG1 and BOSS1244-QG2. BAGPIPES \citep{Carnall2018} enables to fit observed spectroscopic and photometric SEDs with spectral  synthesis models to derive the probability distribution functions (PDFs)  for the SFH, dust and metallicity content of each galaxy. We find that the results measured with BAGPIPES are consistent with the results from FAST++ . In upper panels of Figure~\ref{fig:fige2}, we show the probability distributions of SFRs, mass-weighted age, stellar mass, specific SFR and dust attenuation with 16th, 50th and 84th percentiles marked by the black dashed lines assuming a double power-law SFH model for BOSS1244-QG1 and BOSS1244-QG2. All the results are consistent with the conclusion that BOSS1244-QG1 and BOSS1244-QG2 are massive, quenched galaxies, and  BOSS1244-QG1 is older than BOSS1244-QG2.  Note that only five photometric data points are available, so there are relatively large uncertainties in these parameters. We adopt the results of FAST++ hereafter.

\begin{figure}[!ht]
\centering
\includegraphics[trim=0mm 0mm 0mm 1mm,clip,width=\columnwidth]{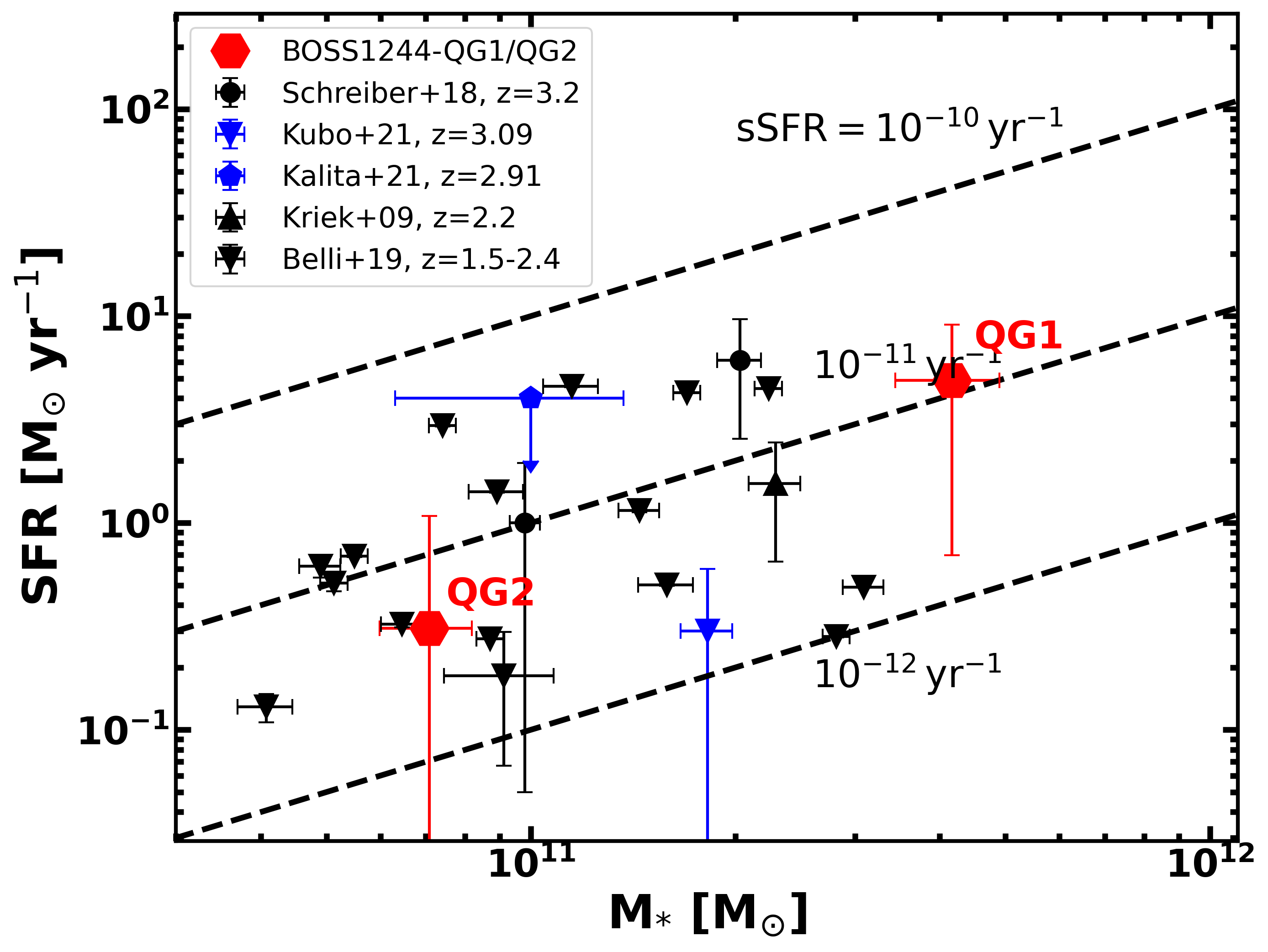}
\caption{The relationship between stellar mass and SFR. The red filled hexagons are BOSS1244-QG1 and BOSS1244-QG2. The blue symbols are the quiescent galaxies in protocluster SSA22 \citep{Kubo2021} at $z=3.09$ and galaxy group RO-1001 \citep{Kalita2021} at $z=2.91$. The black symbols are the quenched galaxies in the random fields \citep{Kriek2009, Schreiber2018, Belli2019}. The dashed lines show the sSFR of 10$^{-10}$\,yr$^{-1}$, 10$^{-11}$\,yr$^{-1}$ and 10$^{-12}$\,yr$^{-1}$.} 
\label{fig:fig6}
\end{figure}

The sSFRs of BOSS1244-QG1 and  BOSS1244-QG2 are log(sSFR)$= -1.93$\,Gyr$^{-1}$ and  log(sSFR)$=-2.36$\,\rm Gyr$^{-1}$, respectively. 
In Figure~\ref{fig:fig6}, we compare stellar mass $M_{*}$  and SFR for quiescent galaxies in the blank field with those in protocluster fields at $z=1.5-3.2$ collected from the literature \citep{Chu2021, Noordeh2021, Zirm2012}, finding no difference on the properties of the quenched galaxies between the blank field and this protocluster.  We conclude that BOSS1244-QG1 and BOSS1244-QG2 are comparable to quiescent galaxies in protoclusters or the fields at $z=2-3$.

\subsection{ BCG formation in  BOSS1244} \label{sec:spatial}

\begin{figure}[!ht]
\centering
\includegraphics[trim=0mm 0mm 0mm 1mm,clip,width=\columnwidth]{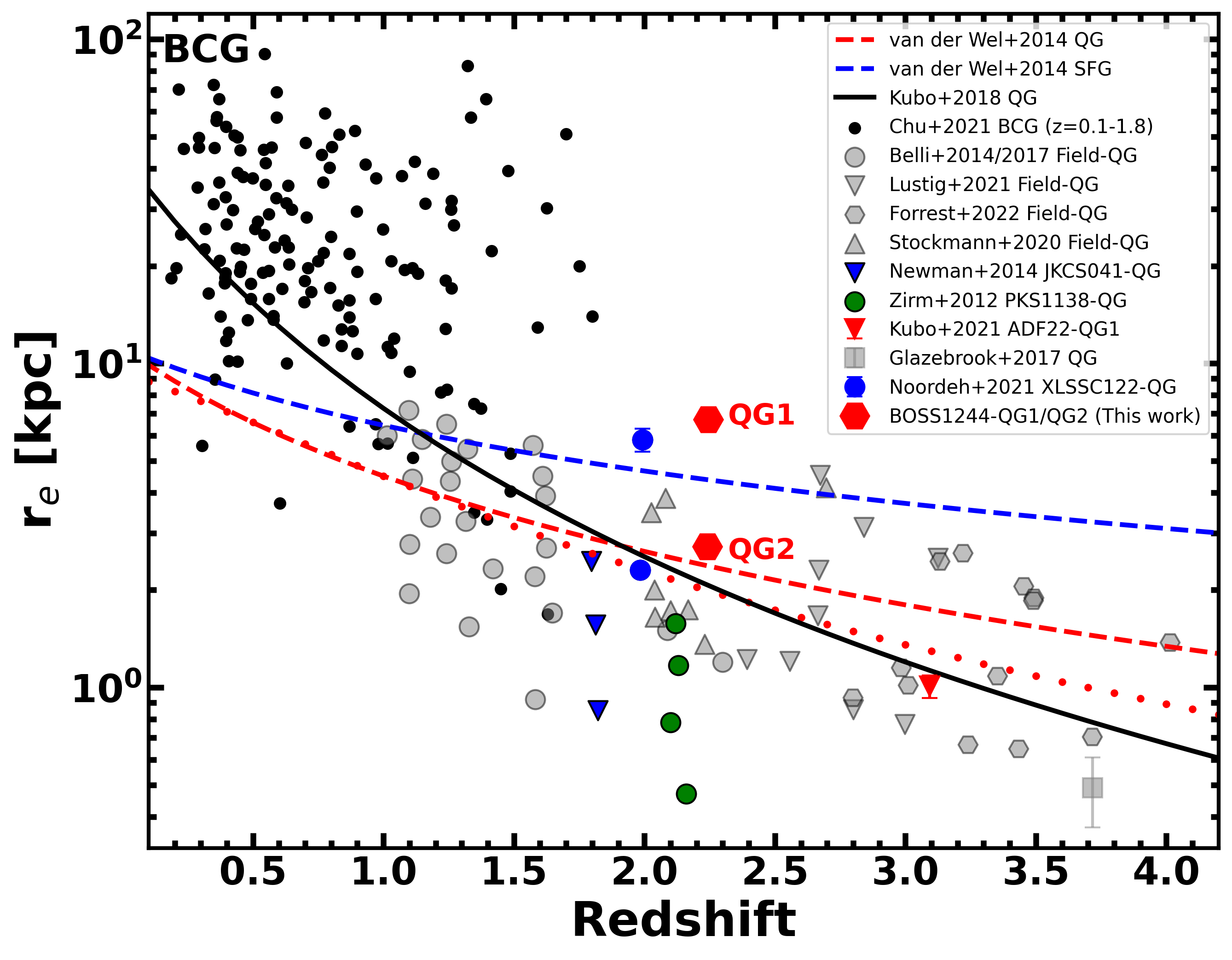}
\caption{The size evolution of massive ($\sim10^{11-11.5}$\,M$_{\odot}$) quiescent or BCG galaxies. The red filled hexagons are BOSS1244-QG1 and BOSS1244-QG2. The black  points are the BCGs \citep{Chu2021} at $z=0.1-1.8$. The color-coded symbols are the quiescent galaxies in (proto)clusters \citep{Newman2014, Zirm2012, Kubo2021, Noordeh2021}, and the grey symbols are the quiescent galaxies in the fields \citep{Belli2014, Belli2017, Glazebrook2017, Stockmann2020, Lustig2021, Forrest2022}. The black and red lines are the size evolution of quiescent galaxies \citep{vanderWel2014,Kubo2018}, the blue line marks the size evolution of star-forming galaxies \citep{vanderWel2014}.} 
\label{fig:fig7}
\end{figure}

In Figure~\ref{fig:fig7}, we collect BCG sample at $z=0.1-1.8$ \citep{Chu2021} and spectroscopically confirmed quiescent galaxies with $M_{*}>10^{11.3}$\,M$_{\odot}$ at $z\sim1-4$ \citep{Zirm2012, Belli2014, Belli2017, Newman2014, Glazebrook2017, Stockmann2020, Lustig2021, Kubo2021, Noordeh2021, Forrest2022}, and present their size evolution. Only a few massive quiescent galaxies are identified in protoclusters (e.g., Spiderweb and SSA22) at $z>2$. Most are found in the general fields, and their large-scale environments are largely unknown. BOSS1244-QG1 and BOSS1244-QG2 follow the size evolution of field early-type galaxies considering the relatively large dispersion, but the size of BOSS1244-QG1 is comparable to the coeval field star-forming galaxies.

The typical S{\'e}rsic index ($n$), effective radius ($r_{\rm e}$) and stellar mass (M$_{*}$) of pure BCGs at $z\sim0$ is $4.45_{-0.11}^{+0.15}$, $11.48_{-0.53}^{+0.79}$\,kpc and $1.95_{-0.09}^{+0.04} \times 10^{11}$\,M$_{\odot}$, while these parameters for their progenitors at $z\sim2$ are predicted  to be $2.32_{-0.34}^{+0.44}$, $3.63_{-0.59}^{+0.25}$\,kpc and $8.13_{-1.12}^{+0.94} \times 10^{10}$\,M$_{\odot}$, respectively \citep{Zhao2017}.   Therefore, these two quiescent galaxies have key characteristics close to those of local BCGs, and are much higher than the parameters expected for progenitors of BCGs at $z\sim2$, especially for BOSS1244-QG1.  

BOSS1244-QG1 and BOSS1244-QG2 already have a high S{\'e}rsic index ($n>4$) and their stellar masses are mostly assembled. Their stellar ages are inferred as $2.14_{-0.19}^{+0.15}$\,Gyr and $0.87_{-0.27}^{+0.02}$\,Gyr, corresponding to a formation redshift of $7.05_{-1.16}^{+1.34}$ and $3.14_{-0.34}^{+0.03}$, respectively. These indicate that BOSS1244-QG2 was formed recently whereas BOSS1244-QG1 had been formed at an earlier time, as shown in the inner panels of Figure~\ref{fig:fige2}. One can conclude that (some) BCGs (e.g., BOSS1244-QG1) have been formed before the virialization of a cluster core.

\section{Discussion} \label{sec:discuss}

\subsection{Average merger timescale of BOSS1244-QG1 and BOSS1244-QG2} \label{sec:spatial}

The projected distance between BOSS1244-QG1 and BOSS1244-QG2 is about $\sim50$\,h$^{-1}$\,kpc. From the HST grism spectra, we identify their redshifts to be  $z=2.2441$ and $z=2.2416$.  At given redshift uncertainties of $\Delta=0.01$, we estimate the velocity separation between the two galaxies is about $231\,$km\,s$^{-1}$. The stellar masses of BOSS1244-QG1 and BOSS1244-QG2 are $>5\times10^{10}$\,M$_{\odot}$. One can suspect that the two galaxies might merge to form a more massive galaxy. If true, this would be the first dry merger ever discovered at $z>2$. It is worthwhile to further explore the possibility that BOSS1244-QG1 and BOSS1244-QG2 are likely in the process of merging. We use the following two methods to estimate the typical merger timescale of BOSS1244-QG1 and BOSS1244-QG2. 

{\bf (a) KW08 model:}   \cite{Kitzbichler2008} derived the average merger timescales for galaxy pairs from the Millennium Simulation. They determined the timescale as functions of redshift and galaxy stellar mass as following
\begin{equation}
\centering
{< T_{{\rm merge}}> }^{-1/2}=T_{0}^{-1/2}+f_{1}z+f_{2}(\rm logM_{*}-10), 
\end{equation}
where the coefficients $T_{0}$, $f_{1}$ and $f_{2}$ and their uncertainties are 3,310\,$h^{-1}$\,Myr, $-1.05\pm$0.03$\times$10$^{-3}$\,$h^{-1}$\,Myr$^{-1/2}$ and 6.68$\pm$0.08$\times$10$^{-3}$\,$h^{-1}$\,Myr$^{-1/2}$ when the radial velocity difference of $\leq300\,$km\,s$^{-1}$ and projection distance of $\leq50$\,h$^{-1}$\,kpc are adopted. 
This  relation depends very little on cosmological parameters and galaxy formation assumptions. Instead, the orbital times of pairs is a function of projected separation and galaxy properties \citep{Kitzbichler2008}. Utilizing this relation, we estimate a mean merger timescale for BOSS1244-QG1 and BOSS1244-QG2 to be $1.41\,$Gyr,  corresponding to an end epoch of the merging at redshift $z\sim1.47$.

 {\bf (b) HLB2022 model:} \cite{Husko2022} studied galaxy merger rates and merger timescales over $0<z<10$ based on galaxy formation models with more accurate tracking of subhalo orbits, giving   
\begin{equation}
\begin{split}
\centering
T_{\rm merge}(M_{*},z,r_{max},v_{max})=&T^{500}_{20}(M_{*},z)\times(\frac{r_{max}}{20\,h^{-1}\,kpc})^{\alpha}\\& \times\frac{erf(v_{max}/V_{0})^{\beta}}{erf(500\,km\,s^{-1}/V_{0})^{\beta}}, 
\end{split}
\end{equation}
and 
\begin{equation}
\centering
T^{500}_{20}(M_{*},z)=T_{0}e^{b_{0}(z-z_{0})^{3}}(\frac{M_{*}}{10^{10}})^{a_{0}+a_{1}(1+z)^{a_{2}}}, 
\end{equation}
where the parameters $T_{0}$, $b_{0}$, $z_{0}$, $a_{0}$, $a_{1}$, $a_{2}$, $V_{0}$, $\alpha$ and $\beta$ are given in table~2 of \cite{Husko2022}. 
We use these formulae with stellar mass and redshift, as well as close projected distance and velocity separation of pairs from  BOSS1244-QG1 and BOSS1244-QG2 and estimate the merger timescale for BOSS1244-QG1 and BOSS1244-QG2 is $1.00\pm0.80\,$Gyr.  This yields an end epoch at redshift $z\sim1.64_{-0.32}^{+0.46}$. The large uncertainty is due to the relative large velocity separation uncertainty between BOSS1244-QG1 and BOSS1244-QG2. The high resolution spectra will improve the uncertainty in the future. This merger timescale is similar to the estimation based on KW08.  We adopt this merger timescale in this work given that this model contains multiple parameters.

We use the spherical collapse model to calculate the evolution of densest region of BOSS1244. There are three specific evolutionary steps based on spherical linear collapse model. The first step is turn-around, when the overdense sphere reaches a maximum radius. The corresponding linear overdensity is $\delta_{\rm ta}\approx1.062$. After the turn-around, the overdense sphere is virialized at half its turn-around radius, and the linear overdensity is $\delta_{\rm vir}\approx1.580$. Then the sphere continues to collapse until the collapse stops, at the linear overdensity of $\delta_{\rm c}\approx1.686$ \citep{Pace2017}. 

\begin{figure}[!ht]
\setlength{\abovecaptionskip}{0pt}
\begin{center}
\includegraphics[trim=0mm 0mm 0mm 1mm,clip,height=0.35\textwidth]{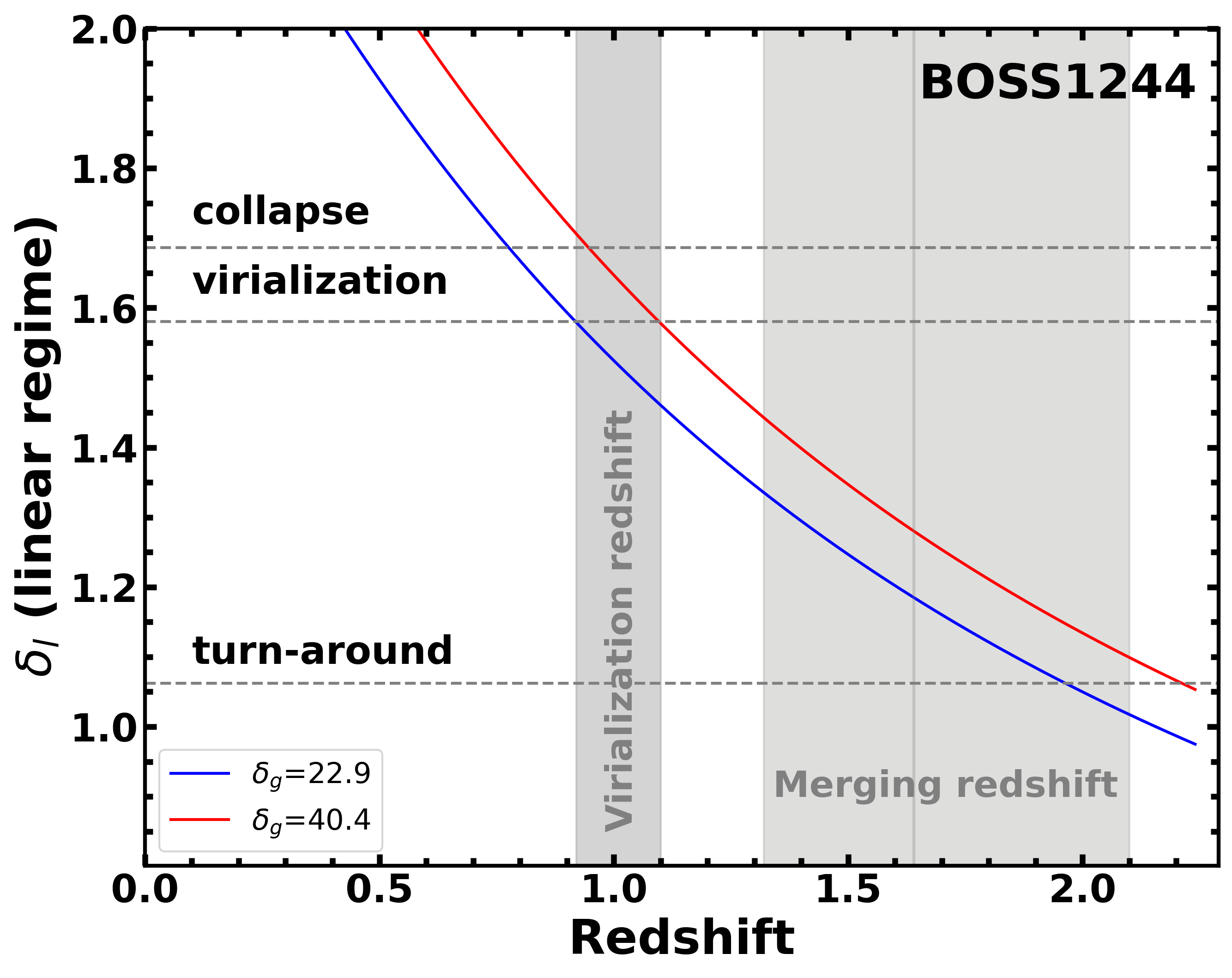}
\caption{Evolution of overdense region in BOSS1244.The evolution is calculated in a linear regime. The blue and red curves are the galaxy overdensity ($\delta_{\rm g}$) of 22.9 and 40.4, respectively. The horizontal dashed lines mark $\delta_{\rm ta}\approx1.062$, $\delta_{\rm vir}\approx1.580$ and $\delta_{\rm c}\approx1.686$. The vertical line is merger redshift of BOSS1244-QG1 and BOSS1244-QG2. A mature BCG though dry major merger could be generated before the virialization of a cluster core.}
\label{fig:fige6}
\end{center}
\end{figure}

The evolution of density peak region in BOSS1244 is calculated using Equation~7 in \cite{Cucciati2018}, and results are  shown in Figure~\ref{fig:fige6}. The blue curve is the evolution of typical protocluster-scale region (galaxy overdensity $\delta_{\rm g}=22.9$ at the scale of 15\,cMpc). The evolution of protocluster core region (galaxy overdensity $\delta_{\rm g}=40.4$ at the scale of 8\,cMpc) is described by the red curve. The definition of galaxy overdensity $\delta_{\rm g}$ is presented in \cite{Shi2021}. Accordingly, the protocluster is expected to become a virialized system by $z\sim0.92$ and the protocluster core  be virialized at $z\sim1.10$, which is later than the end of merging between BOSS1244-QG1 and BOSS1244-QG2. 
Nevertheless,  the collision of BOSS1244-QG1 and BOSS1244-QG2 might end before the virialization of a cluster core would complete. Our results provide direct observational evidence that (some) BCGs can be formed through dry merger in unvirialized core region of a protocluster.  This is subject to the usual expectation for the formation of BCGs in the cluster environments. Here we point out that the adopted merger timescale is one of the largest uncertainty sources, another one is the uncertainty in velocity dispersion derived from the low-resolution of the HST grism slitless spectra. Future high resolution spectroscopic observations (e.g., by JWST) will help to provide improved constraints.

 Although BOSS1244-QG1 and BOSS1244-QG2 follow the stellar mass-size relation of field quiescent galaxies at $z=2-3$ within the relatively large scatter \citep{vanderWel2014,Mowla2019}, they could continue to grow and form a more massive BCG  through dry major merger. Meanwhile, theoretical simulations  demonstrate that the large scatter at the high-mass end of the stellar mass–size relation is driven by dry mergers \citep{Nipoti2003}. Given that dry major merging is essential to form a mature BCG \citep{Bernardi2007, Liu2009, Liu2015, vanDokkum2015}, our observations demonstrate that BOSS1244-QG1 and BOSS1244-QG2, as the progenitors of a mature BCG, have already quenched their star formation before the virialization of the BOSS1244 cluster core. This reveals that the star formation quenching of BCGs and even their progenitors may not be driven by the virialization process of a cluster core at high redshifts. In addition, BCGs usually host AGNs more frequently than coeval field early-type galaxies \citep{Best2007}. AGN feedback heats up the ICM around the protocluster core so that hot gas could emit in X-ray wavelengths \citep{Best2007, VonDerLinden2007}. Such processes could be investigated with X-ray observations in the future.

\subsection{Comparison with of BCGs/QGs in (proto)clusters at $z=0.1-2.0$} \label{sec:core}

\begin{figure*}[!ht]
\setlength{\abovecaptionskip}{0pt}
\begin{center}
\includegraphics[trim=0mm 0mm 0mm 0mm,clip,height=0.35\textwidth]{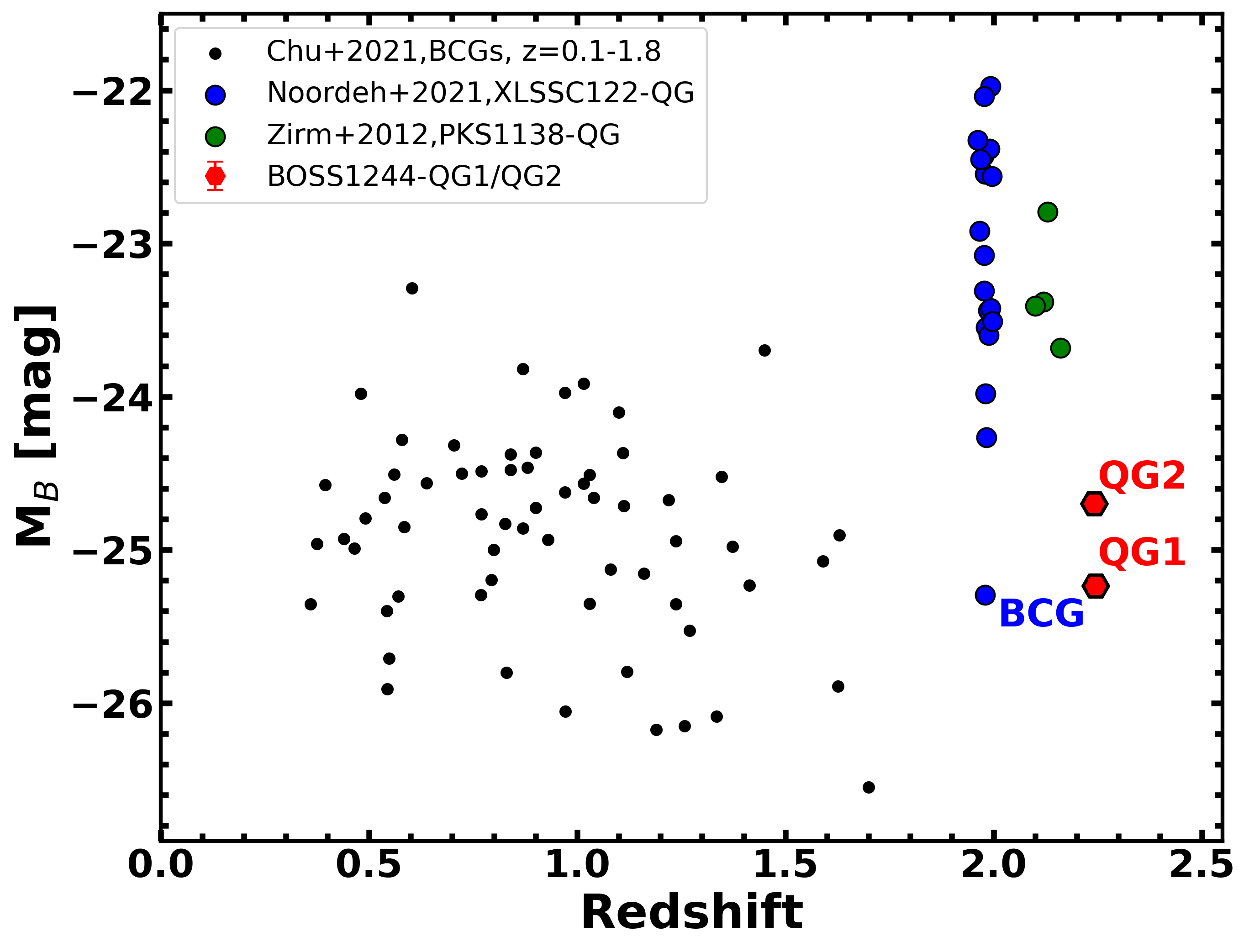}
\includegraphics[trim=0mm 0mm 0mm 0mm,clip,height=0.35\textwidth]{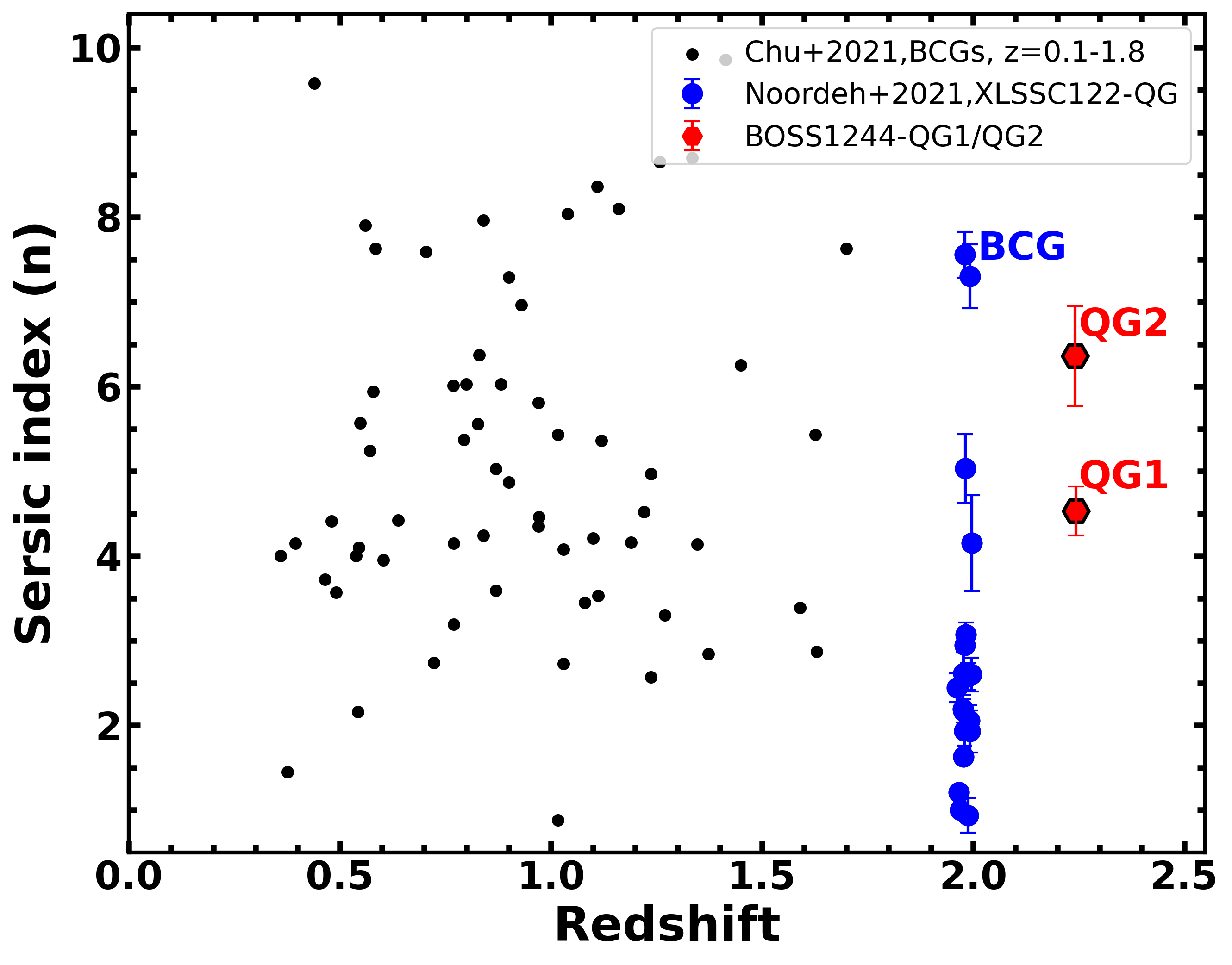}
\caption{Rest-frame $B-$band absolute magnitude (left) and S{\'e}rsic index (right) as a function of redshift. The black points are the BCGs with single component at $z=0.1-1.8$, the blue points are the quiescent galaxies and a marked BCG in the mature cluster XLSSC122 at $z\sim2$ \citep{Chu2021, Noordeh2021}. The green points are the quiescent galaxies selected from protocluster PKS1138 at $z=2.16$ \citep{Zirm2012}. The red filled hexagons are BOSS1244-QG1 and BOSS1244-QG2.} 
\label{fig:fig8}
\end{center}
\end{figure*}

BOSS1244-QG1 and BOSS1244-QG2 are two brightest galaxies in the BOSS1244 SW region. In Figure~\ref{fig:fig8}, we compare BCGs with single component at $z=0.1-1.8$ \citep{Chu2021} and several quiescent galaxies in (proto)cluster at $z\sim2$ \citep{Zirm2012, Noordeh2021}, finding that BOSS1244-QG1 and BOSS1244-QG2 are comparable to local BCGs in terms of their $B$-band absolute magnitude. BOSS1244-QG1 is as bright as the BCG in the mature cluster XLSSC122 at $z\sim2$, and BOSS1244-QG1 and BOSS1244-QG2 are brighter than the quiescent galaxies in the mature cluster XLSSC122 and protocluster PKS1138 at $z=2.16$. The effective radii of BCGs/QGs tend to become larger with time, as shown in Figure~\ref{fig:fig7}. The sizes of BOSS1244-QG1 and BOSS1244-QG2 are smaller than the BCG in the mature cluster XLSSC122, but larger than most quiescent galaxies in XLSSC122 at $z\sim2$ and protocluster PKS1138 at $z=2.16$.  The S{\'e}rsic indices show no evolution with redshift, shown in the right of Figure~\ref{fig:fig8}. The vast majority of BCGs at $z<1.8$ are bulge-like galaxies. The high S{\'e}rsic indices of BOSS1244-QG1 and BOSS1244-QG2 are consistent with the BCGs at $z<2.0$.  

\begin{figure}[!ht]
\setlength{\abovecaptionskip}{0pt}
\begin{center}
\includegraphics[trim=0mm 0mm 0mm 0mm,clip,height=0.35\textwidth]{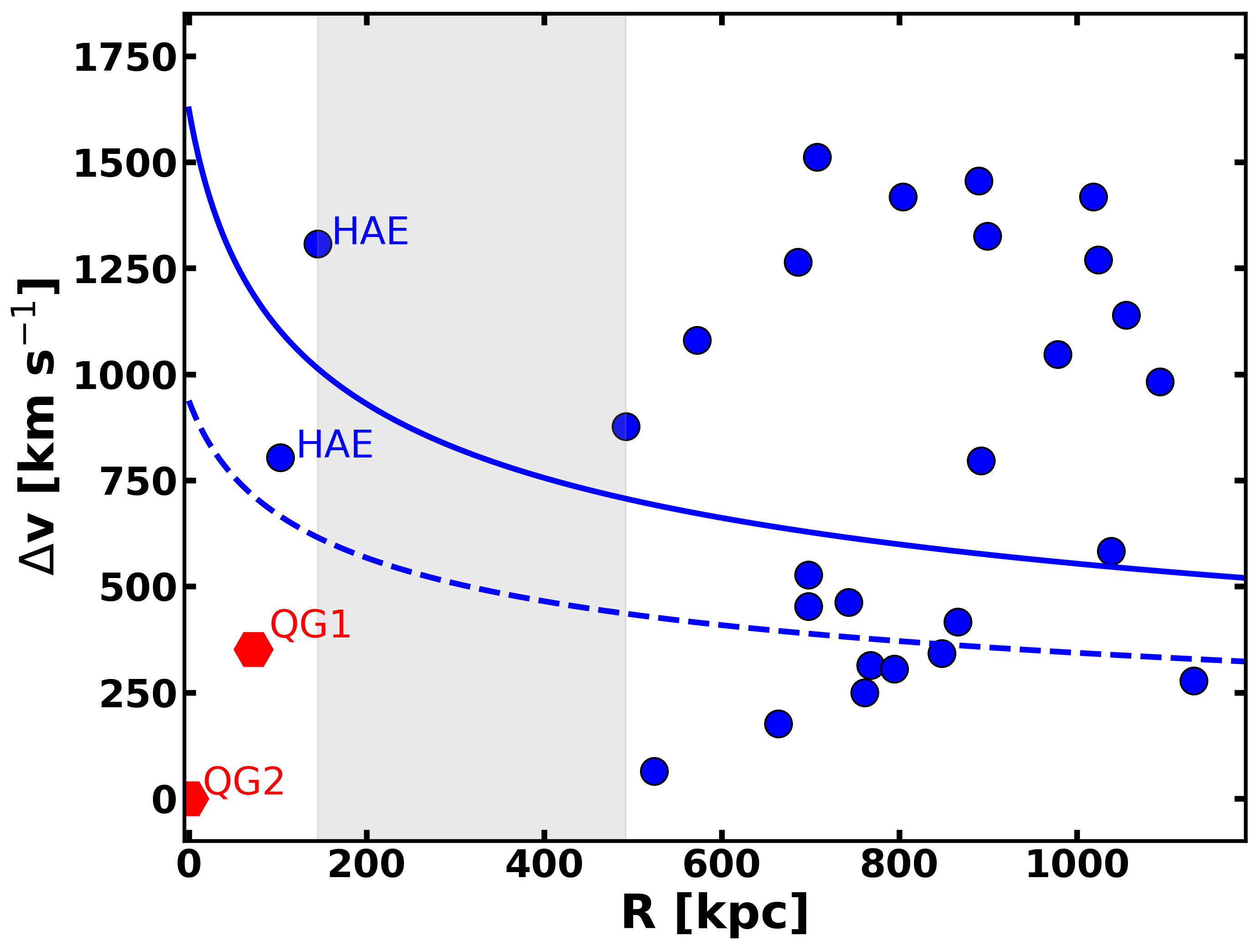}
\caption{The relative velocity and projection distance relative to BOSS1244-QG2.  The red filled hexagons are BOSS1244-QG1 and BOSS1244-QG2 and blue points are the confirmed member galaxies in BOSS1244. The blue solid curve is the escape velocity for an Navarro–Frenk–White (NFW) halo with a halo mass of $10^{13.0}$\,M$_{\odot}$. The blue dashed line is the projection escape velocity (the velocity and distance are divided with $\sqrt{3}$ and $\sqrt{1.5}$ , respectively.).  The grey shaded area show a gap of 347\,kpc.}
\label{fig:fige5}
\end{center}
\end{figure}

Although BOSS1244-QG1 and BOSS1244-QG2 have key characteristics of BCGs, they are likely to grow in size and mass. In Figure~\ref{fig:fige5}, we show the  velocity offset distribution and spatial distances of  the member galaxies relative to BOSS1244-QG2 in BOSS1244.  The  projected separation between BOSS1244-QG1 and BOSS1244-QG2 is 8$\farcs$84, i.e., $\sim$72.8\,kpc at $z=2.24$. Their velocity offset is 231\,km\,s$^{-1}$. Dry mergers between them may promote their increase in size and mass.  Besides, there are two HAEs at  $z=2.250/2.227$ with stellar mass of $10^{10.3}$/$10^{11.1}$\,M$_{\odot}$ around BOSS1244-QG1 and BOSS1244-QG2 within the radial distance of $R\sim145$\,kpc. 
These two HAEs may interact with the two QGs, but the merger timescale between HAEs and QGs is estimated to be larger than 2.0\,Gyr.
As time goes on, the protocluster will become richer and at the same time, BOSS1244-QG1 and BOSS1244-QG2 are expected to undergo a dry merger and create a mature BCG with a dramatically increased size.

In Figure~\ref{fig:fige5}, we find that there is a gap ($\sim$347\,kpc) between central region and  other region for these member galaxies, also see spatial distribution of members in the left panel of Figure~\ref{fig:fig1}. If the gap is true, the core region within 492\,kpc may be affected by accretion shock, so that there is not enough supply of cold gas \citep{Benson2011}.   However, we do not rule out the possibility of being limited by sample completeness, although HST grism with three orbits depth (3$\sigma$ limiting line flux of $1.7\times10^{-17}\,$erg\,s$^{-1}$\,cm$^{-2}$) could scan all the objects in the density peak region of BOSS1244. This phenomenon can be examined with deeper spectroscopic observations and X-ray radiation in the future.

\subsection{Witnessing star formation quenching in protocluster core} \label{sec:core}

From the spatial distribution of BOSS1244 member galaxies in Figure~\ref{fig:fig1}, we notice that the two quiescent galaxies BOSS1244-QG1 and BOSS1244-QG2 reside at the center of the densest region traced by HAEs, normal star-forming galaxies (blue points in Figure~\ref{fig:fig1}) spread in the intermedium densest region, and submilimeter galaxies (SMGs, brown squares in Figure~\ref{fig:fig1}), which are extreme dusty starbursts, are mostly in the outskirts of BOSS1244 \citep{Zhang2022}. The lack of SMGs inside the core region of BOSS1244 hints a transition to the environment disfavoring the extreme starbursts \citep{Zhang2022}. The presence of quiescent galaxies BOSS1244-QG1 and BOSS1244-QG2 at the center and an increase of specific SFR at increasing radial distance suggest an evolutionary picture for the protocluster BOSS1244:  the entire region of  BOSS1244 at early time ($z>3$) grows rapidly; the protocluster core region starts to quench star formation in galaxies, and extends gradually towards the outskirts; still strong star formation activities take place at the periphery of the protocluster.   \cite{Chiang2017} described three phases of the history of cluster formation through a set of $N$-body simulations and semi-analytic models:  galaxy growth in protoclusters proceeded in an inside-out manner at $z=5-10$;  an extended star formation phase at $z=1.5-5$, and a violent infalling and quenching phase at $z=0-1.5$.  At $z=1.5-5$, \cite{Chiang2017}  pointed out that some protocluster cores may be first regions to show evidence of galaxy quenching or dense intracluster gas. Our result supports the theoretical analyzes.

The possible mechanisms for quenching, include internal processes (the feedback from AGN and stellar feedback, and the build-up of hot intracluster gas) \citep{Fabian2012, Man2018}, as well as the external processes (such as ram pressure stripping \citep{Gunn1972, Fumagalli2014, Poggianti2017, Roberts2019}, galaxy-galaxy interactions \citep{Farouki1981}, strangulation \citep{Peng2015} and harassment \citep{Moore1996}). The latter processes mostly happen in the overdense regions.  According to the relation between stellar mass and halo mass, the estimated halo masses of BOSS1244-QG1 and BOSS1244-QG2 are above the critical shock-heating mass of 10$^{12}$\,M$_{\odot}$, the hot halo shocks infalling cold gas from intergalactic medium  to the viral temperature, resulting in shutting down the supply of the fuel for star formation in galaxies \citep{Birnboim2003, Dekel2006, Dekel2009}.  However, the hot halo gas can cool through radiation and condense to form stars after the shock. Additional gas heating from AGN feedback could prevent further the star formation in massive halos \citep{Gabor2010, Man2018}. In addition, galaxy interactions are suggested to transform galaxies, high speed galaxy encounters in dense environments may cause distortions and disk heating, tending to transform disks to ellipticals \citep{Farouki1981, Springel2005, Hopkins2008, Man2018}. The spheroidal morphologies of BOSS1244-QG1 and BOSS1244-QG2 support this possible quenching mechanism, and these two quiescent galaxies are expected to form a more massive quiescent  BCG through major merger. Our NIR spectroscopy of HAEs reveals that the densest region of BOSS1244 contains two substructures \citep{Shi2021}, so the quenching of BOSS1244-QG1 at $z=2.2441$ and BOSS1244-QG2 at $z=2.2416$ in the density peak of BOSS1244 may be affected by cosmological or gravitational heating effects.These two quiescent galaxies quenched star formation before the virialization of the cluster core, revealing that the quenching is not regulated by the processes related to virialization or hot intracluster gas. 
The quenching in BOSS1244-QG2 is likely a faster process than that in BOSS1244-QG1 based on the estimated stellar age. The rapid quenching may result in compact early-type galaxies with high S{\'e}rsic index and the slow quenching, instead, may be attributed to a smooth decline in the gas reservoir, likely caused by fuel starvation \citep{Belli2019}.

\section{Summary} \label{sec:summary}

BOSS1244 with NE and SW components is identified as one of the most massive galaxy protoclusters discovered to date at $z=2-3$. Its densest region is unvirialized and contains two substructrues in redshift space, providing a unique chance to investigate the massive quiescent galaxies. Submillimeter objects detected by JCMT/SCUBA-2 850\,$\micron$ are located in the outskirts of the densest region (SW), and show a prominent offset  from the density peak of HAEs, implying the occurrence of violent star formation enhancement in the outskirts of HAE density peak. Using NIR ground-based and HST WFC3 grism slitless spectroscopy, we identify 40 galaxies at $z=2.223-2.255$ as the member galaxies in the BOSS1244 SW region. The results are summarized as follows. 

\begin{enumerate}[i)]

	\item We discover a pair of massive quiescent galaxies, BOSS1244-QG1 at $z=2.2441$ and BOSS1244-QG2 at $z=2.2416$, residing in the densest region of the BOSS1244 field. BOSS1244-QG1 and  BOSS1244-QG2 are the brightest galaxies in the sample of 40 spectroscopically confirmed member galaxies over $2.223<z<2.255$. The two galaxies are separated by 8$\farcs$84 (72.8\,kpc), and their velocity difference along light of sight is estimated to be $\sim231$\,km\,s$^{-1}$.  

	\item  BOSS1244-QG1 and BOSS1244-QG2 already have a high S{\'e}rsic index ($n>4$) and their stellar masses are sufficiently assembled. Their stellar ages are inferred as $2.14_{-0.19}^{+0.15}$\,Gyr and $0.87_{-0.27}^{+0.02}$\,Gyr, corresponding to a formation redshift of $7.05_{-1.16}^{+1.34}$ and $3.14_{-0.34}^{+0.03}$, respectively. These indicate that BOSS1244-QG2 was quenched recently whereas BOSS1244-QG1 had been quenched at an earlier time. These two quiescent galaxies have key characteristics close to those of local BCGs, and are much higher than the parameters expected for progenitors of BCGs at $z\sim2$, especially for BOSS1244-QG1. One can conclude that (some) BCGs (e.g.,BOSS1244-QG1) have been formed before the virialization of a cluster core. 

	\item BOSS1244-QG1 and BOSS1244-QG2 are expected to merge and form a mature BCG via dry merger. We estimate the merger could end at a timescale of $1.00\pm0.80$\,Gyr, corresponding to the end epoch of the merger at redshift $z\sim1.64_{-0.32}^{+0.46}$, while the protocluster is expected to become virialized  by $z\sim0.92$ and by $z\sim1.10$ for the protocluster core. These suggest that (some) BCGs formed through dry major merger may occur in the unvirialized core region of a protocluster, against the usual expectation for the formation of BCGs in the cluster environment.  

	\item Strikingly, we find that there is a clear gradient in galaxy star formation intensity from the center to the outskirts of BOSS1244, and a strong relationship between EW of  H$\alpha$ emission line and radial distance relative to BOSS1244-QG2 for BOSS1244 member galaxies within a scale of 17.63\,cMpc. This indicates that we are witnessing galaxies in BOSS1244 density peak region are quenching from the center gradually to the outskirts.  

	\item The quenching of these two quiescent galaxies may be affected by gravitational shock heating. AGN feedback cannot be ruled out. These two quiescent galaxies quenched star formation before the virialization of the cluster core, revealing that the quenching is not regulated by the processes related to virialization or hot intracluster gas.

\end{enumerate}

Taken together, our results unveil that formation of massive BCGs through dry major mergers may occur earlier than the full assembly of a cluster core, and the star formation quenching in BCGs and their progenitors can be completed before the cluster core virialization at $z=2-3$. It is clear that the quenching in BCGs is not regulated by the processes related to virialization or hot intracluster gas.  Massive dry pairs are thought to be a good tracer of protocluster cores ($\sim36\arcsec=300\,$kpc in radius at $z=2.24$), especially for the most overdense systems in the early Universe \citep{Ando2020}.  
BOSS1244-QG1 and BOSS1244-QG2 provide direct evidence for the emergence of ultra-massive quiescent galaxies through dry merger in the core region of massive protoclusters. In the future, James Webb Space Telescope (JWST) with unprecedented sensitivity and resolution in imaging and spectroscopy promises a breakthrough in understanding of the formation of BCGs in protoclusters at high redshifts.

\acknowledgments
We thank the anonymous referee for the valuable and constructive comments and suggestions that greatly improved the manuscript. We thank Dr. Yu Luo for his useful discussions in this work. This work is supported by the National Science Foundation of China (12233005, 12073078, 12173088 and 12303015), the science research grants from the China Manned Space Project with NO. CMS-CSST-2021-A02, CMS-CSST-2021-A04 and CMS-CSST-2021-A07.  XW is supported by NASA through HST grants HST-GO-16276, HST-GO-17159, the Fundamental Research Funds for the Central Universities, and the CAS Project for Young Scientists in Basic Research, Grant No. YSBR-062.  D.D.S. acknowledges the supports from China Postdoctoral Science Foundation (2021M703488), Jiangsu Funding Program for Excellent Postdoctoral Talent (2022ZB473) and the National Science Foundation of Jiangsu Province  (BK20231106).  


\clearpage

\bibliographystyle{aasjournal}

\bibliography{ms}{}

\end{document}